\documentclass[aps,onecolumn,prd,showpacs,showkeys,preprintnumbers,superscriptaddress,nobibnotes,notitlepage,floatfix,longbibliography,nofootinbib]{revtex4-2}

\pdfoutput=1

\usepackage{graphicx} 
\usepackage{mathtools}
\usepackage{enumitem}
\usepackage{amssymb,amsmath}
\usepackage{float,xcolor}
\usepackage{soul}
\usepackage{subcaption}
\usepackage{aas_macros}

\def\be{\begin{equation}}
\def\ee{\end{equation}}
\def\bea{\begin{eqnarray}}
\def\eea{\end{eqnarray}}

\def\s{{\rm s}}

\def\cm{{\rm cm}}
\def\km{{\rm km}}
\def\pc{{\rm pc}}
\def\kpc{{\rm kpc}}
\def\mpc{{\rm Mpc}}
\def\gev{{\rm GeV}}

\begin{document}

\title{Scaling Relations for Dark Matter Halos Hosting Ultra-Faint Dwarf Galaxies}

\author{Levi K.~C.~Fisher}
\thanks{{\scriptsize Email}: \href{mailto:jkumar@hawaii.edu}{levikf@hawaii.edu}}
\affiliation{Department of Physics and Astronomy, University of Hawai'i, Honolulu, HI 96822, USA}

\author{Isabelle S.~Goldstein}
\thanks{{\scriptsize Email}: \href{mailto:jkumar@hawaii.edu}{igoldste@monmouth.edu}}
\affiliation{Department of Chemistry and Physics, Monmouth University, 400 Cedar Avenue, West Long Branch, NJ 07764 USA}
\affiliation{Department of Physics and Astronomy,
Mitchell Institute for Fundamental Physics and Astronomy,
Texas A\&M University, College Station, TX  77843, USA}

\author{Jason Kumar}
\thanks{{\scriptsize Email}: \href{mailto:jkumar@hawaii.edu}{jkumar@hawaii.edu}}
\affiliation{Department of Physics and Astronomy, University of Hawai'i, Honolulu, HI 96822, USA}

\author{Louis E.~Strigari}
\thanks{{\scriptsize Email}: \href{mailto:jkumar@hawaii.edu}{strigari@tamu.edu}}
\affiliation{Department of Physics and Astronomy,
Mitchell Institute for Fundamental Physics and Astronomy,
Texas A\&M University, College Station, TX  77843, USA}

\begin{abstract}
We consider the extraction of
parameters of dark matter halos hosting ultra-faint dwarf galaxies,
in the case where there are only ${\cal O}(10)$ identified member stars with measured line-of-sight velocities. This scenario is likely to be increasingly common, as upcoming newly discovered dwarf galaxies in the Milky Way, by e.g. the Rubin Observatory,  will likely (at least initially) have only a few identified members.  Assuming an NFW dark matter profile, equilibrium modeling likely can only robustly extract one halo parameter ($\rho_s r_s$), but the scale radius itself will typically be unconstrained.  In these cases, the results obtainable from Jeans modeling can be well replicated by a simple scaling
relation motivated by the half-light mass estimator.  As an application, we examine the recently discovered stellar system Ursa Major III/Unions 1, which has been optimistically assessed to have the largest $J$-factor of any known object.  We suggest that, because of the presence of outlier stars, the $J$-factor obtained from modeling of Ursa Major III/Unions 1 is likely inflated, as it is inconsistent with the half-light mass estimator, while removal of the outliers will leave the $J$-factor unconstrained from below.
\end{abstract}

\maketitle

\section{Introduction}

Dwarf spheroidal galaxies (dSphs) are key targets for the indirect detection of dark matter~\cite{Fermi2010}.  Because they are dominated by dark matter, they provide an environment in which one can potentially observe gamma-rays arising from dark matter annihilation, free from significant astrophysical backgrounds~\cite{Conrad:2015bsa,Strigari:2018utn}. For this purpose, it is important to estimate the dark matter content of any dSph; the relevant information is encoded in the $J$-factor.  The $J$-factor is typically inferred from the motions of stars within the dSph using, for example, equilibrium Jeans modeling. As many more dSphs are expected to be discovered in upcoming years~\cite{2019arXiv190201055D},
and may provide new pathways in the indirect detection of dark matter, it is important to understand how much
information about the dark matter content of a dSph may be obtained.

The faintest dSphs are the population of ultra-faint dwarf galaxies, which are typically characterized by luminosities in the range hundreds to thousands of times the solar luminosity. In most of these cases, kinematics may be measured for only $\sim$ dozens of stars~\cite{Simon:2007dq,Simon:2019nxf}. While they are promising candidates for dark matter detection~\cite{Strigari:2007at}, the inherent uncertainties in the dark matter distributions provide a systematic that must be accounted for in the analysis.

We focus on the case of these ultrafaint dSphs, and study what may be extracted from their kinematic data. These galaxies are a key benchmark, because they are likely to resemble the systems that will be discovered in the upcoming years. The key question is whether, with
so few identified stars, it is possible not only to determine that the object is a dwarf spheroidal galaxy, but
also to determine all of the parameters of the dark matter halo necessary to estimate the $J$-factor.  Our main
result is that, with only a few dozens stars with available kinematic data, there is likely not enough data
to constrain the scale radius of the halo.  Although one may be able to determine with some certainty the
dark matter density within the half-light radius~\cite{Walker:2009zp}, one likely cannot determine with much certainty where the NFW~\cite{Navarro:1996gj}-predicted $1/r$ dark matter cusp ends, as there are not enough identified stars which probe the gravitational potential at the edge of the cusp.  Indeed, the halo parameters related to the dark matter density in the central regions, which
can be obtained from Jeans modeling, are essentially already constrained by the half-light mass estimator, and can be determined from a simple scaling relation to coarse-grained stellar observables, without the need to resort to detailed Jeans modeling.  One other parameter is then needed to determine the $J$-factor, namely the size of the region over which dark matter annihilation is significant.  This parameter is difficult to determine from Jeans modeling, and rather is set simply by the size of the aperture over which one chooses to view the dSph in an indirect detection search.

As an application of these techniques, we consider Ursa Major III/Unions 1 (UMaIII/U1)~\cite{2024ApJ...961...92S}, which is a recently discovered stellar system that has
garnered significant interest because it is very nearby, and may have a $J$-factor which is potentially larger than that
of any other dSph~\cite{2024ApJ...965...20E}.  But there are significant uncertainties in the $J$-factor estimate of UMaIII/U1.  One key
uncertainty is in the velocity-dispersion of the member stars, which is dominated by the presence of two velocity
outliers, whose membership may be debatable.  We show that including both outliers, or even only one, produces a
half-light mass which is strongly inconsistent with expectations of the half-light mass estimator, indicating the
velocity dispersion is likely overestimated.  But removing both outliers leaves only an upper bound on the velocity
dispersion, which in turn implies only an upper bound on the $J$-factor.  We conclude that there is reason to doubt
that UMaIII/U1's $J$-factor is really as large as the most optimistic estimates, and that this issue can
only be resolved with better stellar data.

The plan of this paper is as follows.  In Section~\ref{sec:Jeans} we briefly review
Jeans modeling, and connect the results of this analysis to a scaling relation obtained from
the half-light mass estimator in the case where there are few member stars.  In Section~\ref{sec:JFactor},
we describe the relationship between what can be determined from Jeans modeling and the $J$-factor.
In Section~\ref{sec:UrsaMajorIII}, we apply these results to Ursa Major III.  We conclude with a
discussion of our results in Section~\ref{sec:DiscussionConclusion}.

\section{Jeans Modeling and the Half-light Mass Estimator}
\label{sec:Jeans}

We begin by briefly reviewing Jeans modeling using the spherical Jeans equation.
The starting assumption of Jeans modeling is that the stellar phase space density $f_\star$ is
spherically symmetric and static.  Liouville's Theorem then implies that
$\{ f_\star , H\} =0$.  Multiplying this equation by the radial momentum ($p_r$) and integrating
over all momenta then yields the spherical Jeans equation,
\bea
\frac{\partial}{\partial r} \left(\rho_\star \langle v_r^2 \rangle \right)
+\rho_\star \left(\frac{\partial \Phi}{\partial r} + \frac{2}{r} \beta_\star \langle v_r^2 \rangle \right) &=& 0 ,
\label{eq:Jeans}
\eea
where $\rho_\star (r)$ is the stellar density profile,
$\langle v_r^2 \rangle (r)$ is the squared stellar radial velocity dispersion, $\Phi (r)$ is the
gravitational potential, and $\beta_\star (r) \equiv 1 -\langle v_\perp^2 \rangle / 2  \langle v_r^2 \rangle$
is the spherical anisotropy.  Given an ansatz for $\rho_\star$, $\beta_\star$ and the dark matter density
profile ($\rho_{DM}$), eq.~\ref{eq:Jeans} is a first order ordinary differential equation for $\langle v_r^2 \rangle (r)$,
which can be solved exactly (see, for example,~\cite{Goldstein:2022pxu}).
Projecting the velocity dispersion onto the line-of-sight (which is the easiest
quantity to measure observationally), yields a solution for the squared line-of-sight velocity dispersion, $\sigma_{los}^2 (R)$,
where $R$ is the 2d-projected distance from the center of the dSph.
Given this solution, one can find the likelihood of the observed stellar data, which consists of the radial distance $R$ and
line-of-sight velocity $v_{los}$ for each identified member star.  By varying the ansatz for the dark matter density and
marginalizing over the other parameters, such as the stellar anisotropy, one can maximize the likelihood of the data in a Bayesian analysis, the details and prior ranges for which are discussed in Appendix \ref{app:jeansdetails}.
This process yields the posterior on the dark matter halo parameters.

To illustrate how well this process can actually constrain dark matter parameters, we will consider the case in
which the dark matter density profile is taken be of NFW form:
\bea
\rho_{DM} (r) &=& \frac{\rho_s}{ (r/r_s)(1+(r/r_s))^2} .
\eea
This profile depends on two dimensionful parameters, a scale density $\rho_s$ and a scale radius $r_s$.
Note that the dark matter density need not have such a cuspy profile.  In particular, it is believed that
self-interacting dark matter (SIDM) models will tend to yield cored density profiles \cite{2022arXiv220710638A}.  Jeans modeling, when
the profile parameterization allows both cuspy or cored profiles, can generally be consistent with either
form.  However, numerical simulations of halos with non-interacting dark matter tend to yield cuspy profiles which are broadly consistent with the NFW form. Further, we do not account for the fact that subhalos tend to be tidally-stripped and more exponentially truncated in their outer regions. Indeed, given the resolution of simulations at this stage, the nature of the dark matter halos that host the ultra-faint dwarf galaxies is uncertain~\cite{Bullock:2017xww}. For simplicity, we will
assume a density profile of NFW form.

\begin{figure}
    \centering
    \includegraphics[width=0.9\linewidth]{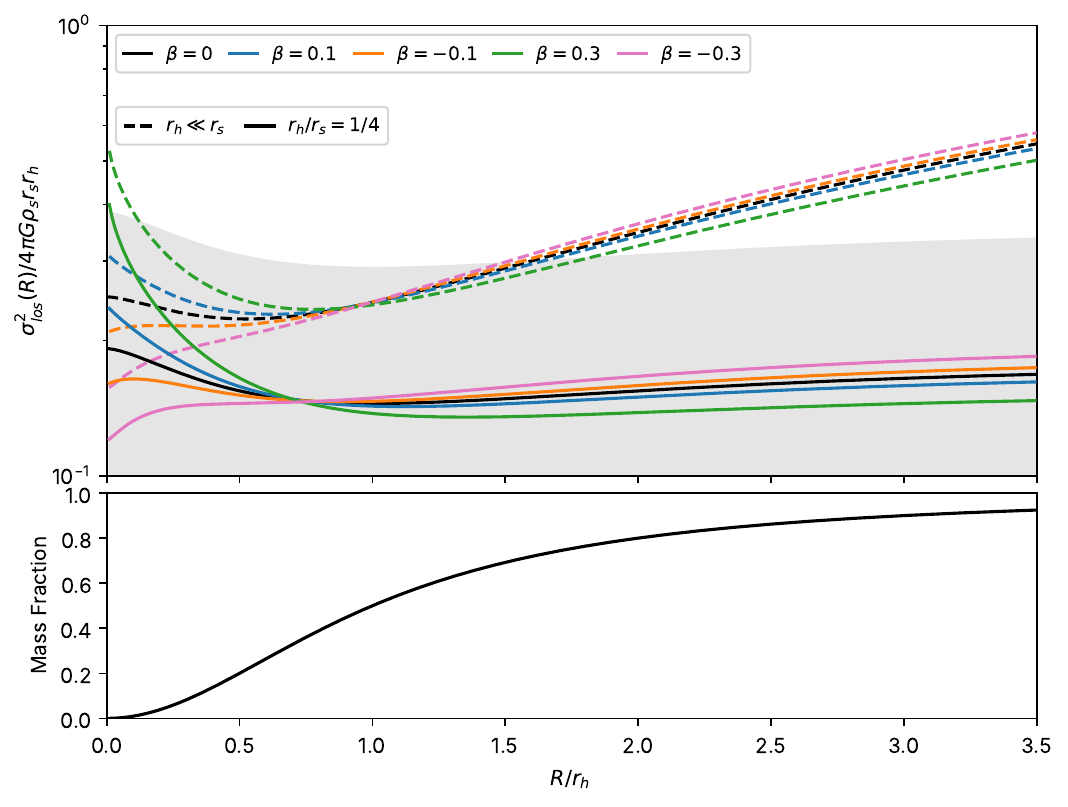}
    \caption{The top panel shows a plot of $\sigma_{los}^2$ as a function of $R/r_h$, the 2D projected radius divided by the azimuthally averaged half-light radius, in units of $4\pi G_N \rho_s r_s r_h$ for various
    choices of $r_h / r_s$ and $\beta_\star$. We assume an NFW dark matter profile, a Plummer stellar profile, and that the gravitational potential is dominated by dark matter.
    The gray band identifies the region $< 2\sigma_{los}^2 (R)$ for
    $r_h / r_s =1/4$, $\beta_\star =0$.  As discussed in the text, if $v_{los}$ is Gaussian-distributed with
    variance $\sigma_{los}^2$, $85\%$ of stars would like in the grey band for this parameter choice.
    The bottom panel shows the fraction of the stellar mass enclosed within $R/r_h$.}
    \label{fig:SigmaLosSq}
\end{figure}

In Figure~\ref{fig:SigmaLosSq}, we plot $\sigma_{los}^2 (R)$ obtained from solving the spherical Jeans equation,
assuming the stellar density profile is of the Plummer form \cite{1911MNRAS..71..460P}: $\rho_\star (r) = \rho_*(0) (1 + (r/r_h)^2)^{-5/2}$.
We assume the dark matter profile is of the NFW form, and that the gravitational potential is dominated by
dark matter (as would be expected for ultrafaint dwarf galaxies, with few observed stars).  Each curve in
Figure~\ref{fig:SigmaLosSq} is labeled by the choice of $\beta_\star$ (assumed constant and indicated by color, as labeled) and of $r_h / r_s$
(indicated by line style, as labeled).
$\sigma_{los}^2 (R)$ is plotted in units of $4\pi G_N \rho_s r_s r_h$, where we hold $r_h$ and $\rho_s r_s$ fixed,
while varying $r_h/r_s$.  Note, this choice of unit allows one to compare on the same footing curves with the
same half-light radius and the same dark
matter profile in the inner slope region ($\rho(r) \sim \rho_s r_s /r$), but with different choices of $r_s$.
We see from Figure~\ref{fig:SigmaLosSq} that it is essentially impossible to distinguish the case $r_h/r_s = 1/4$ from the case $r_h /r_s \ll 1$ unless one observes stars at a 2D projected radial distance $R$ significantly greater than  $r_h$. However, practically searching for member stars at large radius is subject to complications with foreground contamination~\cite{2024MNRAS.527.4209J,2024AJ....167...57T}, and small number statistics.

To illustrate this issue concretely,
we can make a simple assumption that stars at a distance $R$ have a line-of-sight velocity $v_{los}(R)$ drawn from a Gaussian distribution with zero mean and variance $\sigma_{los}^2 (R)$.  If $\beta_\star = 0$, $r_h / r_s = 1/4$, then $85\%$ of stars would then be expected to have $v_{los}^2$ which lies within the light gray band.
The overlap of this band with the dashed curves (representing $r_h / r_s \ll 1$, for various choices of $\beta_\star$)
is indicative of the difficulty in determining $r_h / r_s $ if one only observes stars are small $R$ (especially
since $\beta_\star$ is unknown).
One would need $R > 4 r_h$ in
order to find $\sigma_{los}^2 (R; r_h/r_s \ll 1) / \sigma_{los}^2 (R; r_h/r_s =1/4 ) > 3$ for any $\beta_\star$.
A single star
found at projected distance $R > 4r_h$ could then yield a Bayes factor large enough for the posterior distribution
to substantially favor one value of $r_h/r_s$ over the other.
But for a Plummer profile, only $\sim 5\%$ of stars would be expected to have a radial distance $R > 4r_h$.
One would thus suspect that, if there are fewer than 20 identified member stars with available kinematics,
the posterior distribution of $r_h/r_s$ would be prior-dominated, with Jeans modeling providing no real
constraint on $r_s$.

\begin{figure*}
\begin{subfigure}{0.9\textwidth}
    \includegraphics[width=\textwidth]{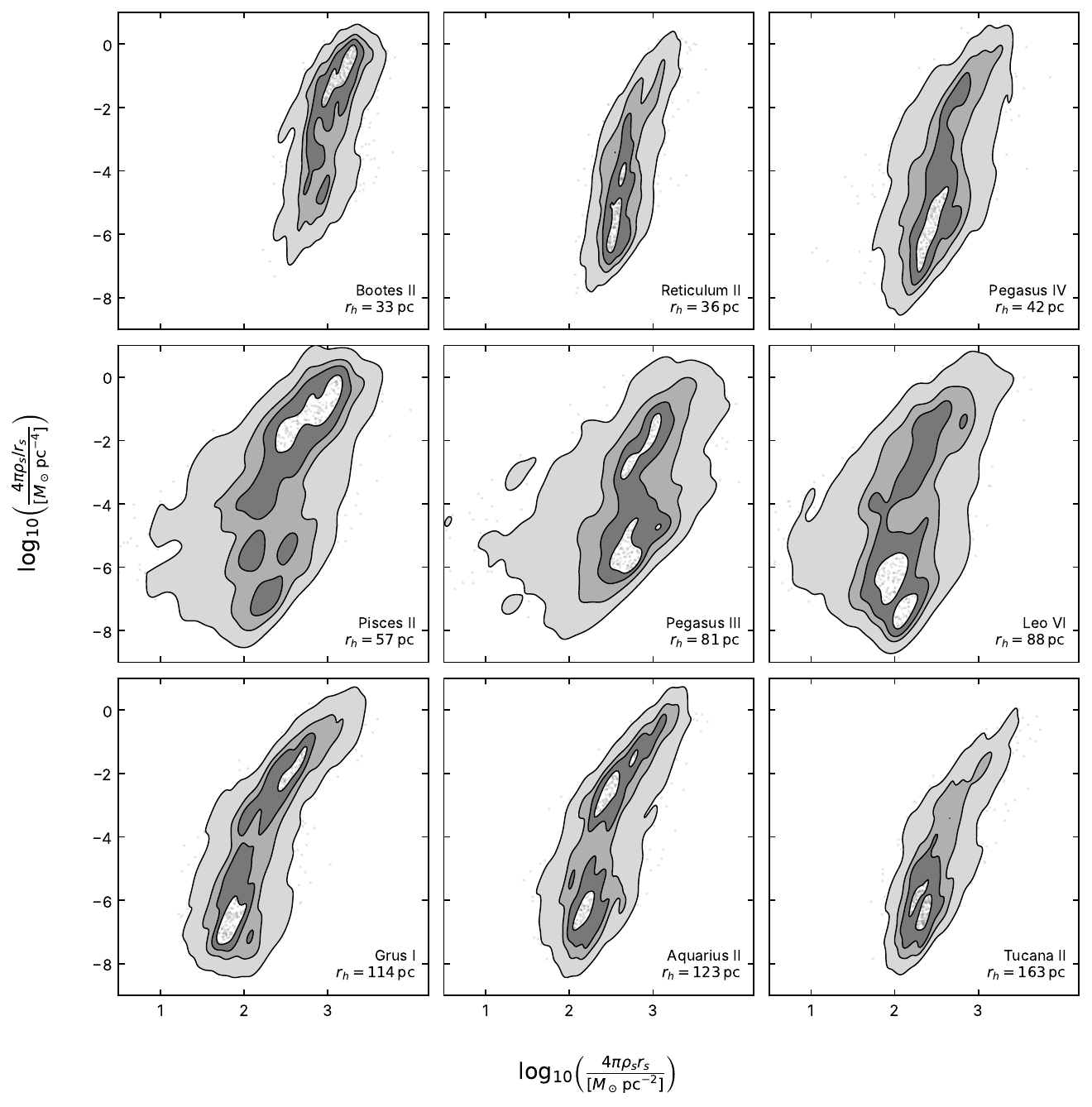}
\end{subfigure}
\caption{The $\rho_s / r_s$ and $\rho_s r_s$ posteriors for 9 dSphs (labeled).
The contours, from darker to lighter, show the $(0.5, 1, 1.5, 2)\sigma$ 2D-Gaussian-equivalent credible intervals.
}
\label{fig:app_rhosrs}
\end{figure*}

We can confirm this intuition by performing Jeans modeling for 9 dSphs for which the number of identified member stars
with measured line-of-sight velocities is less than 20.
For details of the Jeans analysis, see Appendix~\ref{app:jeansdetails}.
In particular,\footnote{Membership is determined by the groups cited here.}
we consider Aquarius II~\cite{Bruce_2023},
Bootes II~\cite{Bruce_2023}, Grus I~\cite{Chiti_2022}, Leo VI~\cite{Tan_2025},
Pegasus III~\cite{2016ApJ...833...16K}, Pegasus IV~\cite{2023ApJ...942..111C},
Pisces II~\cite{2015ApJ...810...56K}, Reticulum II~\cite{2015ApJ...811...62K} and Tucana II~\cite{Walker_2016}.
For each, we perform Jeans modeling assuming an NFW
dark matter profile, Plummer stellar profile, and constant stellar anisotropy $\beta_\star$, using stellar kinematic data
obtained from the references above.
We plot the $\rho_s / r_s$ and $\rho_s r_s$ posteriors for the 9 dSphs we have considered  in Figure~\ref{fig:app_rhosrs}.
Note that there is
significant degeneracy between $\rho_s$ and $r_s$, neither of which are well-constrained individually.  But, as expected,
the product $\rho_s r_s$ is well-constrained (to within an order of magnitude), while the ratio $\rho_s / r_s$ is much less
constrained.

\begin{figure}
    \centering
    \includegraphics[width=0.9\linewidth]{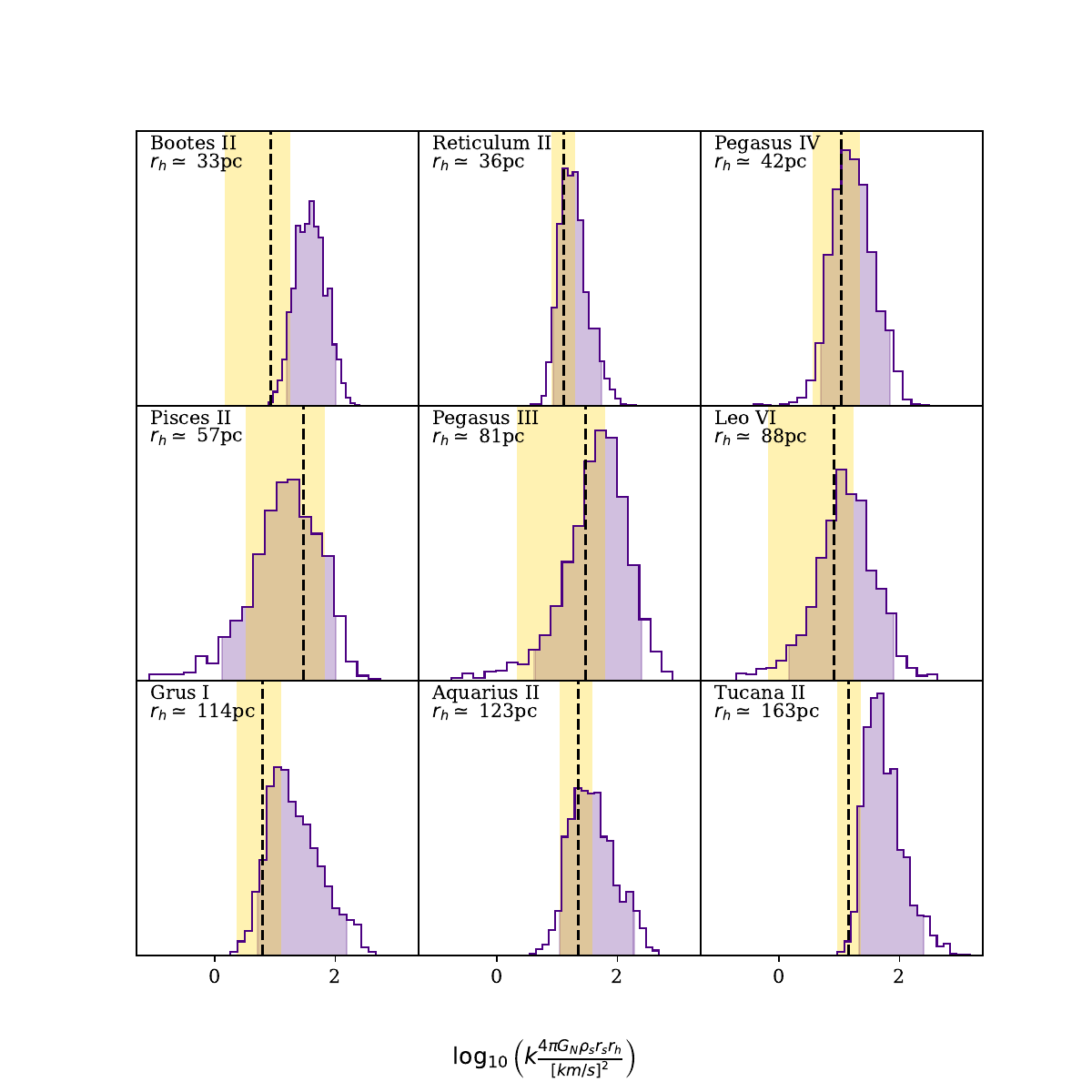}
    \caption{Normalized posterior probability densities for $\log_{10}(4\pi G_N k \rho_s r_s r_h / (\km / \s)^2)$, for 9 dSphs (labeled).
    The purple shaded regions are the $90\%$ containment regions.  Also shown are the mean values of
    $\overline{\sigma}_{los}^2 / (\km / \s)^2$ (dashed black line) and uncertainties (yellow band) found
    in Ref.~\cite{Pace:2024sys}, plotted in the same logarithmic scale.
    }
    \label{fig:rhos_rs}
\end{figure}

To see if the results we obtain from Jeans modeling are consistent with $r_h / r_s \ll 1$, we may use
the half-light mass estimator~\cite{Walker:2009zp,Wolf:2009tu}.  In particular, it was shown that a dark matter
halo will roughly obey the relation
\bea
M_{1/2} &\sim& \frac{5}{2} \frac{\overline{\sigma}_{los}^2 r_{h,3d}}{G_N} ,
\label{eq:HalfLight}
\eea
where $r_{h,3d}$, the 3D-deprojected half-light radius, is the radius of a sphere containing half of the
stellar mass, and $M_{1/2}$ is the total mass contained within that sphere.  Importantly, it was shown that
eq.~\ref{eq:HalfLight} is largely insensitive to the stellar density profile, dark matter density profile, and the
stellar anisotropy, with variations in these parameters yielding only ${\cal O}(1)$ variations in the coefficient
of the half-light mass estimator relation.

For a Plummer stellar profile, we have $r_{h,3d} \sim 1.3~r_h$, where $r_h$ is the 2D-projected half-light radius.
For an NFW dark matter profile satisfying $r_h / r_s \ll 1$, we find $M_{1/2} \sim 2\pi \rho_s r_s r_{h,3d}^2$.  Eq.~\ref{eq:HalfLight} can
then be expressed as
\be
4\pi G_N \rho_s r_s r_h k
\sim \overline{\sigma}_{los}^2 ,
\label{eq:Scaling}
\ee
where $k = 1.3 / 5$.
The posterior distributions of most of the individual dSphs we considered are consistent with the result in
eq.~\ref{eq:Scaling}, as illustrated in Fig.~\ref{fig:rhos_rs}.
The posteriors obtained from Jeans modeling thus seems to robustly indicate consistency with
$r_h / r_s \ll 1$, regardless of assumptions about the stellar anisotropy.

More generally, however, these results indicate that, when there are only ${\cal O}(10)$ identified member stars
with kinematics, it seems likely that the only parameter of the dark matter halo which one can reconstruct
is the product $\rho_s r_s$, rather than either $\rho_s$ or $r_s$ individually.  The scale radius itself cannot be determined from Jeans modeling, since it is unlikely
that there are enough stars at far enough distances to determine the point at which the dark matter density starts to
fall of as $r^{-3}$.  But to determine the parameter $\rho_s r_s$, Jeans modeling is not necessary.  The result one
obtains from Jeans modeling is consistent with a simple scaling relation obtained from the half-light mass-estimator
(eq.~\ref{eq:Scaling}).  Note that, there is generally also a degeneracy between $\rho_s$ and the stellar
anisotropy $\beta_\star$~\cite{1987MNRAS.224...13D,1987ApJ...313..121M,2009MNRAS.393L..50E,Richardson:2012ig,Diakogiannis_2014}, which one might worry would further cloud one's ability to reconstruct the halo parameters.  But
the half-light mass estimator is largely robust to variations in $\beta_\star$~\cite{Walker:2009zp,Wolf:2009tu}, and
for an NFW profile, $\rho_s r_s$ can be
directly determined from $M_{1/2}$.
With so few stars, there is little information to be gained  from Jeans modeling about the halo
parameters beyond this
scaling relation.

\section{Impact on determining the $J$-factor}
\label{sec:JFactor}
The $J$-factor, which is the factor which encodes the dependence of the photon flux produced by dark matter annihilation on the astrophysics of the halo, may be derived from the solutions to the Jeans equations. Expressing the dark matter annihilation cross section as $\sigma_A v = (\sigma_A v)_0 (v/c)^n$ (where $v$ is the relative velocity), then the photon flux arriving from a region of solid angle $\Delta \Omega$ around a dSph is given by
\bea
\Phi_\gamma (\Delta \Omega) &\propto& J_n (\Delta \Omega) ,
\eea
where (see, for example,~\cite{Boddy:2017vpe,Boddy:2019wfg})
\bea
J_n (\Delta \Omega) &=& \int_{\Delta \Omega} d\Omega \int d\ell
\int d^3 v_1 \int d^3 v_2 ~
f(\vec{r}, \vec{v}_1)~ f(\vec{r}, \vec{v}_2)~ (|\vec{v}_1 - \vec{v}_2|/c)^n ,
\eea
where $f(\vec{r},\vec{v})$ is the dark matter phase space distribution, $\vec{\ell}$ is a vector
along the line of sight ($\ell = |\vec{\ell}|$), and $\vec{r} = \vec{\ell} - \vec{D}$, where
$D = |\vec{D}|$ is the distance to the dSph.  The most commonly studied scenario is $n=0$ (velocity-independent
dark matter annihilation), in which case this relation simplifies to
$J_0 (\Delta \Omega) = \int_{\Delta \Omega} d\Omega \int d\ell ~\rho^2 (r)$, where
$\rho(r) = \int d^3 v ~f(r,v)$.

As we have seen in the previous section, with kinematic data from only ${\cal O}(10)$ identified member stars,
the only information which one can reconstruct about the dark matter halo is the quantity $\rho_s r_s$, where we have assumed an NFW dark matter profile. The quantity $\rho_s r_s$ determines the behavior of the dark matter density profile in the inner $1/r$ cusp region. However, this is not sufficient to determine the total $J$-factor, integrated over the
entire halo.  The piece of information which is missing is the scale radius, after which the density falls of $\propto r^{-3}$, and dark matter annihilation becomes negligible.  The scale radius is not well-determined
because there are not enough identified stars at distances large enough to probe the gravitational potential past the probable location of the scale radius.

The additional parameter needed to determine the $J$-factor is related to the observations; specifically, it is the size of the region over which a particular observation will measure gamma-rays arising from dark matter annihilation. This is related to the point spread function of a given experiment. We can denote the radius associated with this region as $r_{end}$.
We should find $r_{end} \lesssim r_s$, since there will certainly be negligible
dark matter annihilation in the region $r \gg r_s$, as noted above. But $r_{end}$ is also constrained by the size of the aperture used in any
particular search for dark matter annihilation in a dSph.  Typical searches for dark matter annihilation in dSphs using Fermi-LAT data assume a point-like source for the annihilation, broadened by the point-spread-function, which is $\lesssim 1^\circ$ for energies of interest for typical dark matter searches. If the scale radius is sufficiently large, we would find $r_{end} \sim D \theta_{aper} \ll r_s$, where $D$ is the distance to the dSph.

The scaling of the $J$-factor with the parameters $\rho_s r_s$ and $r_{end}$ can be determined from dimensional analysis, keeping in mind that the $J$-factor will scale as $D^{-2}$, that is, $\int_{\Delta \Omega} d\Omega \int d\ell ~\rho^2
\sim (1/D^2) \int_{\Delta \Omega} dV~\rho^2$.
For the case of velocity-independent dark matter annihilation, by simple dimensional analysis we then have $J_0 \propto (\rho_s r_s)^2 r_{end} D^{-2}$, where $r_{end} = D \theta_{aper}$. Of course, observations spanning an angle $\theta_{aper}$
will be sensitive to dark matter annihilation along the entire line of sight, including positions whose distance from the center of
the dSph are $> D \theta_{aper}$. This emission is particularly important for dSph in the region of the sky towards the Galactic center, such as Sagittarius~\cite{Vienneau:2024xie}. Thus, in addition to a dependence on $\rho_s r_s$ and $r_{end}$, determined above from dimensional analysis, there will be a weak residual dependence on $r_s$.
We can illustrate this scaling relation and determine the proportionality constant, as well as bound the residual dependence on $r_s$, using two limiting approximations.

First, consider the approximation in which we assume that the dark matter density is given by $\rho(r) = \rho_s r_s / r$ for
$r < r_{end} \ll r_s \ll D$, but there is no dark matter annihilation for $r > r_{end}$.  This is a conservative approximation, because it
correctly models dark matter annihilation within the sphere of radius $r_{end} \ll r_s$, but ignores all dark matter annihilation
outside of this sphere along the line-of-sight.  For this assumption, we can write the $J$-factor as
\bea
J_0 &>& \frac{4\pi}{D^2} \int_0^{r_{end}} dr~r^2 \left(\frac{\rho_s r_s}{r} \right)^2 = 4\pi \frac{(\rho_s r_s)^2 r_{end}}{D^2} .
\label{eq:J_underestimate}
\eea

An alternative limit would be to take $r_s \rightarrow \infty$, which would amount to assuming
$\rho(r) = \rho_s r_s / r$, and integrating along the entire volume encompassed by an aperture of size
$\theta_{aper} = r_{end}/D \ll 1$ along the line-of-sight.
The assumption $r_s \rightarrow \infty$ necessarily leads to an overestimate of the $J$-factor, since we
are ignoring the fact that, at sufficiently large distance along the line-of-sight, the density will fall off faster than $1/r$.
Using the results of Ref.~\cite{Boddy:2019wfg},
we find
\bea
J_0 &<&  \frac{4\pi \rho_s^2 r_s^3}{D^2}
\int_0^{r_{end}/r_s} d\tilde \theta ~ \tilde \theta
\int_{\tilde \theta}^\infty d\tilde r
\left[1 - \left(\frac{\tilde \theta}{\tilde r} \right)^2 \right]^{-1/2} \tilde r^{-2}  = 2\pi^2 \frac{(\rho_s r_s)^2 r_{end}}{D^2} .
\label{eq:J_overestimate}
\eea
We thus find a conservative estimate
\bea
J_0  &=& 4\pi \frac{(\rho_s r_s)^2 r_{end}}{D^2} = 4\pi \frac{(\rho_s r_s)^2 \theta_{aper}}{D} .
\label{eq:J_conservative}
\eea
The conservative underestimate in eq.~\ref{eq:J_underestimate} and the overestimate in eq.~\ref{eq:J_overestimate} differ by a factor of
$\pi/2$, so the uncertainty from residual dependence on $r_s$ is at most $\sim 50\%$.
Using the half-light mass estimator relation (eq.~\ref{eq:Scaling}), we find the
scaling relation
which is applicable when the scale radius encompasses the aperture size (see also~\cite{Evans:2016xwx,Pace:2018tin,Hoskinson:2024hpk}),
\bea
J_0 &=&  4\pi \left(\frac{3.85}{4\pi G_N} \frac{\overline{\sigma}_{los}^2}{r_h} \right)^2 \frac{ \theta_{aper}}{D} ,
\nonumber\\
&=& (10^{A}~\gev^2 \cm^{-5}) \left(\frac{\sigma_{los}}{5~\km/\s} \right)^4
\left(\frac{r_h}{100~\pc} \right)^{-2}
\left(\frac{D}{100~\kpc} \right)^{-1}
\left(\frac{\theta}{1^\circ}\right) ,
\label{eq:scaling}
\eea
where $A$ is a numerical exponent.  Using the nominal
coefficient of the half-light mass estimator would yield $A = 18.49$.

\begin{figure}
    \centering
    \includegraphics[width=0.75\linewidth]{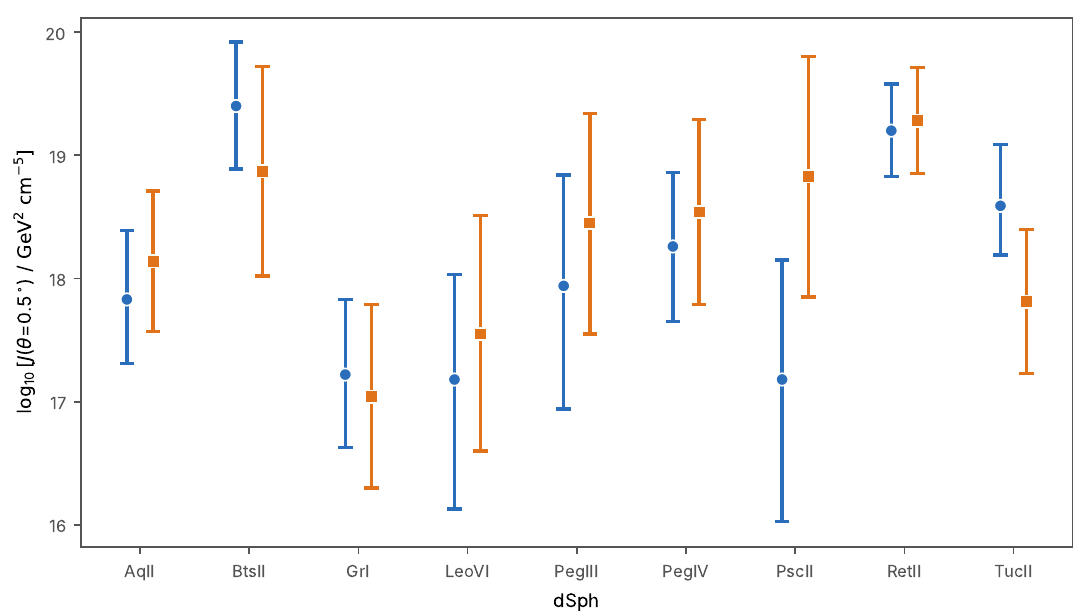}
    \caption{$J$-factors integrated to $0.5^\circ$ for the dSphs we consider.  The blue circles
    are obtained from the posterior distribution, while the orange squares are obtained from
    the scaling relation.}
    \label{fig:Jfactor}
\end{figure}

But the coefficient of the half-light mass estimator is only accurate
up to ${\cal O}(1)$ factors, which depend on the dark matter profile, stellar profile, and
stellar anisotropy.  To account for this we vary $A$ in order to find the value which best fits
the posterior distribution of $J = 2\pi \int_0^{\theta} d\theta'~\sin \theta' \int_0^\infty d\ell~
[\rho(r)]^2$, where $r = \sqrt{D^2 + \ell^2 - 2D\ell \cos \theta'}$.  Doing so, we obtain
the best-fit value $A = 18.76$.
We plot the $J$-factors obtained from eq.~\ref{eq:scaling} (setting $A=18.76$) for each of these
9 dSphs (orange boxes) in Figure~\ref{fig:Jfactor}, setting $\theta = 0.5^\circ$.  The uncertainties in
$J$ are obtained by propagating in quadrature the symmetrized uncertainties in $r_h$, $\sigma_{los}$ and
$D$ reported in Ref.~\cite{Pace:2024sys}.  We also plot the median (blue circles) and 68\% credible
uncertainties of the posterior distribution for $J$.  We see that the scaling relation largely
reproduces the $J$-factors obtained from Jeans modeling.

Note that we have focused thus far on the $J$-factor appropriate for velocity-independent dark
matter annihilation ($J_0$).  But in many scenarios, the dark matter annihilation cross section scales with the
relative velocity as $(v/c)^n$.  In this case, the velocity-dependent $J$-factor
will have additional factors of $(v_0/c)^n$, where $v_0$ is a velocity scale of dark matter within the
region bounded by $r_{end}$.
Determination of the velocity-dependent $J$-factor requires knowledge of the velocity-distribution.
But if one assumes isotropy, then the velocity-distribution is a function of a single integral of motion, the
energy, and can be determined from the density distribution using the method of Eddington inversion~\cite{10.1093/mnras/76.7.572}.  In particular,
the velocity distribution is then determined by the same dimensionful parameters as the density
distribution, $\rho_s$ and $r_s$, and the scaling of the $J$-factor with these parameters is determined as before
by dimensional analysis, up to the additional factor of $(v_0/c)^n$.
Given the parameters $\rho_s r_s$ and $r_{end}$, there is only one
factor one can form with units of velocity, $v_0 = \sqrt{4 \pi G_N (\rho_s r_s) r_{end}}$.
Thus, the scaling relation for velocity-dependent factor, $J_n$, will have an additional factor of $(v_0/c)^n$,
where $v_0$ can be expressed in terms of $r_{end}$ and $\overline{\sigma}_{los}^2 / r_h$ using eq.~\ref{eq:Scaling}
(see also Ref.~\cite{Hoskinson:2024hpk}).
The numerical coefficient can similarly be obtained by explicit integration, using the velocity distribution
obtained from Eddington inversion of an NFW profile, following the procedure of Ref.~\cite{Boddy:2019wfg,Boucher:2021mii}.

\section{An Application to Ursa Major III/Unions 1}
\label{sec:UrsaMajorIII}

We can apply the analysis considered here to the case of Ursa Major III/Unions 1~\cite{2024ApJ...961...92S}, a recently discovered stellar system with a velocity dispersion that may indicate that it has a $J$-factor larger than that of any known dSph.  However, there is significant uncertainty due to the presence
of two velocity outliers among the 11 identified member stars.
Including all 11 identified member stars,
the $J_0$-factor obtained from Jeans modeling is~\cite{2024ApJ...965...20E}
\bea
\log_{10} (J_0 / \gev^2 \cm^{-5}) &=& 21_{-2}^{+1} ,
\eea
and the average line-of-sight velocity dispersion is given by
$\overline{\sigma}_{los} = 3.7_{-1.0}^{+1.4}~\km /\s$.
Removing the largest velocity outlier reduces the velocity dispersion to
$\overline{\sigma}_{los} = 1.9_{-1.1}^{+1.4}~\km /\s$,
which is nearly consistent with zero,
while also removing
the second velocity outlier leaves one only with an upper bound on $\overline{\sigma}_{los}$~\cite{2024ApJ...961...92S}.
To assess how well one can trust the results of Jeans modeling, given these uncertainties,
we treat UMaIII/U1 separately.

\begin{figure*}
    \includegraphics[width=\textwidth]{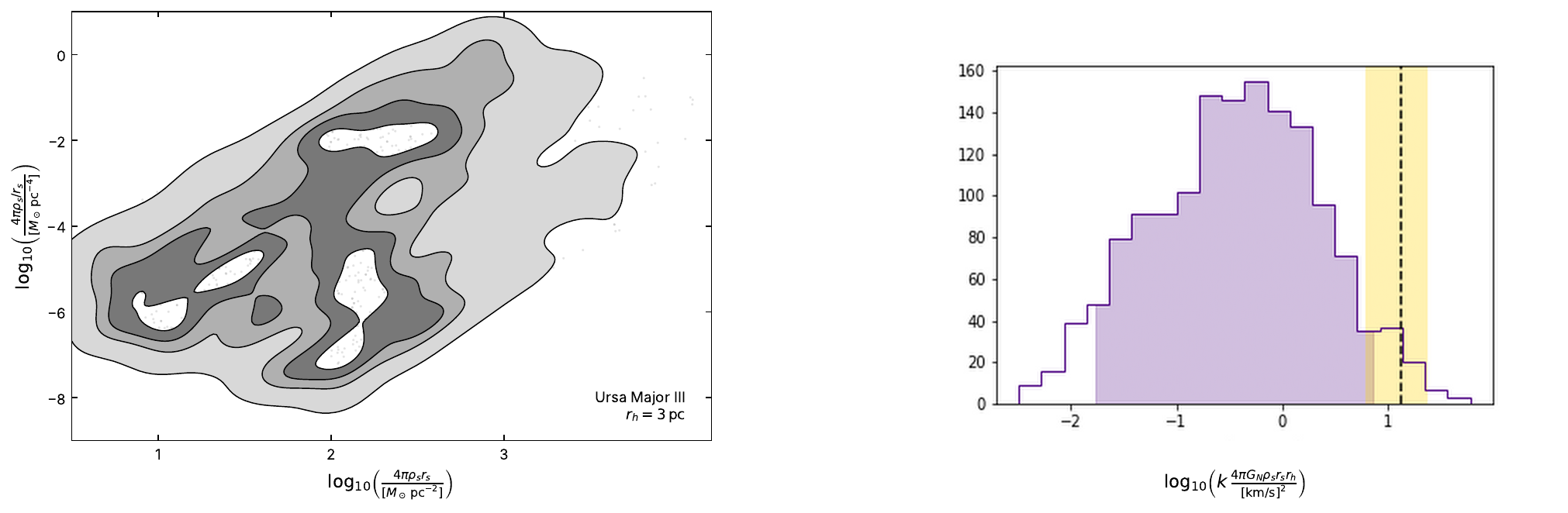}
    \caption{The left panel shows $\rho_s / r_s$ and $\rho_s r_s$ posteriors for Ursa Major III.  The right panel shows the histogram of $\log_{10}(4\pi G_N k \rho_s r_s r_h / (\km / \s)^2)$ for Ursa Major III.
    The purple shaded region is the $90\%$ containment region.  Also shown is the mean value of
    $\overline{\sigma}_{los}^2 / (\km / \s)^2$ (dashed black line) and uncertainties (yellow band) found
    in Ref.~\cite{2024ApJ...961...92S}.  Both panels use
    all 11 member stars identified in Ref.~\cite{2024ApJ...961...92S}.
    }
    \label{fig:UM3}
\end{figure*}

We plot $\rho_s r_s$ and $\rho_s / r_s$ posteriors for UMaIII/U1 (using all 11 member stars) in the left
panel of Figure~\ref{fig:UM3}.  We see that the posterior distribution of $\rho_s r_s$ extends over more than three orders of
magnitude, indicating that this quantity is much less constrained by spherical Jeans modeling for the case of UMaIII/U1 than
for the other dSphs which we have considered.

We plot the posterior distribution for $\log_{10} (4\pi G_N \rho_s r_s r_h k / (\km / \s)^2)$ in
right panel  of Figure~\ref{fig:UM3}, for the case in which all 11 member stars are used.  We also plot $\overline{\sigma}_{los}^2 / (\km / \s)^2$
(dashed black line) and the uncertainties (yellow band) listed above.  The posterior distribution obtained
from Jeans modeling is discrepant with the half-light mass estimator under the assumption $r_h \ll r_s$, despite the very wide range of
the posterior distribution (spanning more than 3 orders of magnitude).  In fact, measured value of $\overline{\sigma}_{los}^2 / (\km / \s)^2$
exceeds the expected value from the posterior distribution by more than an order of magnitude.

We can consider the trustworthiness of the results of Jeans modeling by examining the
half-light mass estimator, assuming a stellar Plummer profile and an NFW dark matter profile, but making no
assumptions about $r_h / r_s$.  If one
performs Jeans modeling with all 11 member stars (using the positions and line-of-sight velocities reported
in~\cite{2024ApJ...961...92S}), one finds
\be
\frac{M_{1/2} G_N}{(5/2) (1.3 r_h)} =
0.38_{-0.34}^{+1.71} ~(\km / \s)^2 .
\label{eq:UrsaMajor3_11}
\ee
But using the half-light mass estimator relation in eq.~\ref{eq:HalfLight} (with $r_{h,3d} = 1.3 r_h$), we
see that the posterior on $M_{1/2}$ is inconsistent with expectations ($M_{1/2}G_N /
(5/2) r_{h,3d} \sim \overline{\sigma}_{los}^2 = 13.7~(\km / \s)^2$).
This might lead one to suspect that the line-of-sight velocity
dispersion was being overestimated by inclusion of velocity outliers.

But as we see from eq.~\ref{eq:scaling}, reducing $\overline{\sigma}_{los}$ will reduce $J_0$.  This suggests that
the $J$-factor obtained for UMaIII/U1 from Jeans modeling with 11 member stars overestimates the $J$-factor,
and also that one cannot obtain a lower bound on $J_0$ without obtaining more precise stellar velocity dispersion data, in order to find a lower bound on $\overline{\sigma}_{los}$ consistent with the half-light mass estimator.

After this work was completed, a new study was presented~\cite{2026ApJ...999L...8C} analyzing the kinematics and metallicity of
UMaIII/U1 members, using updated observations.  This analysis now found 16 member stars, but one of them was
confirmed as a binary, with three others found to be candidate binaries.  In particular, the one confirmed binary was a
velocity outlier in the analysis of Ref.~\cite{2024ApJ...961...92S}.  A binary complicates the determination of the
line-of-sight velocity dispersion, since it is difficult to disentangle the motion of binaries around the dwarf galaxy center of
mass from the motion of each individual star about the binary center of mass.  Removing the binaries, the analysis of Ref.~\cite{2026ApJ...999L...8C}
found only an upper bound on the line-of-sight velocity dispersion ($\overline{\sigma}_{los} < 2.3~\km / \s$  at $95\%$ CL), concluding that
there is no positive evidence for the presence of dark matter in UMaIII/U1.  This conclusion was strengthened by an analysis
of the metallicity of member stars, which found a low metallicity and small metallicity dispersion that is more consistent with a star cluster
than a dwarf galaxy.  We may conclude that the results of Ref.~\cite{2026ApJ...999L...8C} are consistent with our own analysis.

\section{Discussion and Conclusion}
\label{sec:DiscussionConclusion}

Several new dSphs have been discovered in recent years, and it is expected that the advent of the Vera Rubin
Observatory will only increase the pace.  Any new dSph becomes an interesting target for dark matter indirect
detection searches, especially if, as in the case of the recently discovered Ursa Major III/Unions 1,
it is relatively nearby.
It thus becomes important to extract the parameters of the host dark matter halo, in order to determine the
halo $J$-factor.

One would typically expect newly discovered dSphs to initially have
only a few identified member stars with measured line-of-sight velocities, since dSphs with many easily identified
member stars would have been discovered long ago.  This brings up the question of how well one can extract the
parameters of the halo, when there are only a few stars to act as tracers of the potential.

We have argued that the scale radius of the halo is likely to be unconstrained if there are only ${\cal O}(10)$ identified
stars, as there will not be enough stars to probe the region of the gravitational potential where the density falls off
dramatically.  The one parameter which can be measured through Jeans modeling, assuming an NFW profile, is $\rho_s r_s$,
and the result will be largely degenerate with what is obtained using the half-light mass estimator.  This result can
be expressed in terms of a simple scaling relation to coarse-grained stellar observables.  But determining the $J$-factor
requires another parameter, which essentially determines how large is the region in which the dark matter annihilation
rate is large.  When this region is large compared to the stellar half-light radius, the boundary of the annihilation
region is essentially determined by the aperture size of the observation.  In particular, for the case of
velocity-independent ($s$-wave) dark matter annihilation, we find $J_0 \propto \theta_{aper}$.

This leads to the question of whether one can expect to improve sensitivity to dark matter annihilation simply by
increasing the aperture size.  We can see that increasing the aperture likely will not dramatically improve
sensitivity.  Although the number of photons observed arising from dark matter annihilation will scale as
$S \propto J_0 \propto \theta_{aper}$, the number of events due to astrophysical fore/backgrounds will scale as
the solid angle, $B \propto \theta_{aper}^2$.  The signal significance, $S / B^{1/2}$, will thus be largely independent
of the aperture size.

As an application of these results, we considered the case of UMIII/U1, a recently discovered nearby stellar system,
which is optimistically estimated to have the largest known $J$-factor.  However, there are questions regarding how
much one can trust the $J$-factor obtained for UMIII/U1 via Jeans modeling, given that two of the 11 identified
member stars may be velocity outliers.  We find that the half-light mass obtained from Jeans modeling of
UMIII/U1 is inconsistent with the half-light mass estimator, possibly indicating that the velocity-outliers have biased
the average line-of-sight velocity dispersion upward, yielding an overestimate for the $J$-factor.  Removing the outliers
can yield consistency with the half-light mass estimator, as the line-of-sight velocity dispersion then only has an
upper bound, implying only an upper bound on the $J$-factor.  It appears then that the robust determination of the
$J$-factor of UMIII/U1 may require improved stellar data.

{\bf Acknowledgements}  For facilitating portions of this research, ISG, JK and LES wish to acknowledge the Center for Theoretical Underground Physics and Related Areas (CETUP*), the Institute for Underground Science at Sanford Underground Research Facility (SURF), and the South Dakota Science and Technology Authority for hospitality and financial support, as well as for providing a stimulating environment.
The technical support and advanced computing resources from University of Hawai'i Information Technology Services – Research Cyberinfrastructure, funded in part by the National Science Foundation CC* awards \#2201428 and \#2232862, are gratefully acknowledged.
JK is supported in part by DOE grant DE-SC0010504.
LES is supported in part by DOE grant DE-SC0010813.
The authors acknowledge the use of Anthropic Claude to assist with figure production and some statistical
analysis.  The authors have verified all such results and take responsibility for their accuracy.

\bibliography{thebib.bib}

\begin{thebibliography}{43}%
\makeatletter
\providecommand \@ifxundefined [1]{%
 \@ifx{#1\undefined}
}%
\providecommand \@ifnum [1]{%
 \ifnum #1\expandafter \@firstoftwo
 \else \expandafter \@secondoftwo
 \fi
}%
\providecommand \@ifx [1]{%
 \ifx #1\expandafter \@firstoftwo
 \else \expandafter \@secondoftwo
 \fi
}%
\providecommand \natexlab [1]{#1}%
\providecommand \enquote  [1]{``#1''}%
\providecommand \bibnamefont  [1]{#1}%
\providecommand \bibfnamefont [1]{#1}%
\providecommand \citenamefont [1]{#1}%
\providecommand \href@noop [0]{\@secondoftwo}%
\providecommand \href [0]{\begingroup \@sanitize@url \@href}%
\providecommand \@href[1]{\@@startlink{#1}\@@href}%
\providecommand \@@href[1]{\endgroup#1\@@endlink}%
\providecommand \@sanitize@url [0]{\catcode `\\12\catcode `\$12\catcode
  `\&12\catcode `\#12\catcode `\^12\catcode `\_12\catcode `\%12\relax}%
\providecommand \@@startlink[1]{}%
\providecommand \@@endlink[0]{}%
\providecommand \url  [0]{\begingroup\@sanitize@url \@url }%
\providecommand \@url [1]{\endgroup\@href {#1}{\urlprefix }}%
\providecommand \urlprefix  [0]{URL }%
\providecommand \Eprint [0]{\href }%
\providecommand \doibase [0]{https://doi.org/}%
\providecommand \selectlanguage [0]{\@gobble}%
\providecommand \bibinfo  [0]{\@secondoftwo}%
\providecommand \bibfield  [0]{\@secondoftwo}%
\providecommand \translation [1]{[#1]}%
\providecommand \BibitemOpen [0]{}%
\providecommand \bibitemStop [0]{}%
\providecommand \bibitemNoStop [0]{.\EOS\space}%
\providecommand \EOS [0]{\spacefactor3000\relax}%
\providecommand \BibitemShut  [1]{\csname bibitem#1\endcsname}%
\let\auto@bib@innerbib\@empty
\bibitem [{\citenamefont {{Abdo}}\ \emph {et~al.}(2010)\citenamefont {{Abdo}},
  \citenamefont {{Ackermann}}, \citenamefont {{Ajello}}, \citenamefont
  {{Baldini}}, \citenamefont {{Ballet}}, \citenamefont {{Barbiellini}},
  \citenamefont {{Bastieri}}, \citenamefont {{Bellazzini}}, \citenamefont
  {{Blandford}}, \citenamefont {{Bloom}}, \citenamefont {{Bonamente}},
  \citenamefont {{Borgland}}, \citenamefont {{Bouvier}}, \citenamefont
  {{Brandt}}, \citenamefont {{Bregeon}}, \citenamefont {{Brigida}},
  \citenamefont {{Bruel}}, \citenamefont {{Buehler}}, \citenamefont {{Buson}},
  \citenamefont {{Caliandro}}, \citenamefont {{Cameron}}, \citenamefont
  {{Caraveo}}, \citenamefont {{Carrigan}}, \citenamefont {{Casandjian}},
  \citenamefont {{Charles}}, \citenamefont {{Chaty}}, \citenamefont
  {{Chekhtman}}, \citenamefont {{Cheung}}, \citenamefont {{Chiang}},
  \citenamefont {{Ciprini}}, \citenamefont {{Claus}}, \citenamefont
  {{Cohen-Tanugi}}, \citenamefont {{Conrad}}, \citenamefont {{Decesar}},
  \citenamefont {{Dermer}}, \citenamefont {{de Palma}}, \citenamefont
  {{Digel}}, \citenamefont {{Silva}}, \citenamefont {{Drell}}, \citenamefont
  {{Dubois}}, \citenamefont {{Dumora}}, \citenamefont {{Favuzzi}},
  \citenamefont {{Fortin}}, \citenamefont {{Frailis}}, \citenamefont
  {{Fukazawa}}, \citenamefont {{Fusco}}, \citenamefont {{Gargano}},
  \citenamefont {{Gasparrini}}, \citenamefont {{Gehrels}}, \citenamefont
  {{Germani}}, \citenamefont {{Giglietto}}, \citenamefont {{Giordano}},
  \citenamefont {{Glanzman}}, \citenamefont {{Godfrey}}, \citenamefont
  {{Grenier}}, \citenamefont {{Grondin}}, \citenamefont {{Grove}},
  \citenamefont {{Guillemot}}, \citenamefont {{Guiriec}}, \citenamefont
  {{Hadasch}}, \citenamefont {{Harding}}, \citenamefont {{Hays}}, \citenamefont
  {{Jean}}, \citenamefont {{J{\'o}hannesson}}, \citenamefont {{Johnson}},
  \citenamefont {{Johnson}}, \citenamefont {{Kamae}}, \citenamefont
  {{Katagiri}}, \citenamefont {{Kataoka}}, \citenamefont {{Kerr}},
  \citenamefont {{Kn{\"o}dlseder}}, \citenamefont {{Kuss}}, \citenamefont
  {{Lande}}, \citenamefont {{Latronico}}, \citenamefont {{Lee}}, \citenamefont
  {{Lemoine-Goumard}}, \citenamefont {{Llena Garde}}, \citenamefont {{Longo}},
  \citenamefont {{Loparco}}, \citenamefont {{Lovellette}}, \citenamefont
  {{Lubrano}}, \citenamefont {{Makeev}}, \citenamefont {{Mazziotta}},
  \citenamefont {{Michelson}}, \citenamefont {{Mitthumsiri}}, \citenamefont
  {{Mizuno}}, \citenamefont {{Monte}}, \citenamefont {{Monzani}}, \citenamefont
  {{Morselli}}, \citenamefont {{Moskalenko}}, \citenamefont {{Murgia}},
  \citenamefont {{Naumann-Godo}}, \citenamefont {{Nolan}}, \citenamefont
  {{Norris}}, \citenamefont {{Nuss}}, \citenamefont {{Ohsugi}}, \citenamefont
  {{Omodei}}, \citenamefont {{Orlando}}, \citenamefont {{Ormes}}, \citenamefont
  {{Pancrazi}}, \citenamefont {{Parent}}, \citenamefont {{Pepe}}, \citenamefont
  {{Pesce-Rollins}}, \citenamefont {{Piron}}, \citenamefont {{Porter}},
  \citenamefont {{Rain{\`o}}}, \citenamefont {{Rando}}, \citenamefont
  {{Reimer}}, \citenamefont {{Reimer}}, \citenamefont {{Reposeur}},
  \citenamefont {{Ripken}}, \citenamefont {{Romani}}, \citenamefont {{Roth}},
  \citenamefont {{Sadrozinski}}, \citenamefont {{Saz Parkinson}}, \citenamefont
  {{Sgr{\`o}}}, \citenamefont {{Siskind}}, \citenamefont {{Smith}},
  \citenamefont {{Spinelli}}, \citenamefont {{Strickman}}, \citenamefont
  {{Suson}}, \citenamefont {{Takahashi}}, \citenamefont {{Takahashi}},
  \citenamefont {{Tanaka}}, \citenamefont {{Thayer}}, \citenamefont {{Thayer}},
  \citenamefont {{Tibaldo}}, \citenamefont {{Torres}}, \citenamefont {{Tosti}},
  \citenamefont {{Tramacere}}, \citenamefont {{Uchiyama}}, \citenamefont
  {{Usher}}, \citenamefont {{Vasileiou}}, \citenamefont {{Venter}},
  \citenamefont {{Vilchez}}, \citenamefont {{Vitale}}, \citenamefont {{Waite}},
  \citenamefont {{Wang}}, \citenamefont {{Webb}}, \citenamefont {{Winer}},
  \citenamefont {{Yang}}, \citenamefont {{Ylinen}}, \citenamefont {{Ziegler}},\
  and\ \citenamefont {{Fermi LAT Collaboration}}}]{Fermi2010}%
  \BibitemOpen
  \bibfield  {author} {\bibinfo {author} {\bibfnamefont {A.~A.}\ \bibnamefont
  {{Abdo}}}, \bibinfo {author} {\bibfnamefont {M.}~\bibnamefont {{Ackermann}}},
  \bibinfo {author} {\bibfnamefont {M.}~\bibnamefont {{Ajello}}}, \bibinfo
  {author} {\bibfnamefont {L.}~\bibnamefont {{Baldini}}}, \bibinfo {author}
  {\bibfnamefont {J.}~\bibnamefont {{Ballet}}}, \bibinfo {author}
  {\bibfnamefont {G.}~\bibnamefont {{Barbiellini}}}, \bibinfo {author}
  {\bibfnamefont {D.}~\bibnamefont {{Bastieri}}}, \bibinfo {author}
  {\bibfnamefont {R.}~\bibnamefont {{Bellazzini}}}, \bibinfo {author}
  {\bibfnamefont {R.~D.}\ \bibnamefont {{Blandford}}}, \bibinfo {author}
  {\bibfnamefont {E.~D.}\ \bibnamefont {{Bloom}}}, \bibinfo {author}
  {\bibfnamefont {E.}~\bibnamefont {{Bonamente}}}, \bibinfo {author}
  {\bibfnamefont {A.~W.}\ \bibnamefont {{Borgland}}}, \bibinfo {author}
  {\bibfnamefont {A.}~\bibnamefont {{Bouvier}}}, \bibinfo {author}
  {\bibfnamefont {T.~J.}\ \bibnamefont {{Brandt}}}, \bibinfo {author}
  {\bibfnamefont {J.}~\bibnamefont {{Bregeon}}}, \bibinfo {author}
  {\bibfnamefont {M.}~\bibnamefont {{Brigida}}}, \bibinfo {author}
  {\bibfnamefont {P.}~\bibnamefont {{Bruel}}}, \bibinfo {author} {\bibfnamefont
  {R.}~\bibnamefont {{Buehler}}}, \bibinfo {author} {\bibfnamefont
  {S.}~\bibnamefont {{Buson}}}, \bibinfo {author} {\bibfnamefont {G.~A.}\
  \bibnamefont {{Caliandro}}}, \bibinfo {author} {\bibfnamefont {R.~A.}\
  \bibnamefont {{Cameron}}}, \bibinfo {author} {\bibfnamefont {P.~A.}\
  \bibnamefont {{Caraveo}}}, \bibinfo {author} {\bibfnamefont {S.}~\bibnamefont
  {{Carrigan}}}, \bibinfo {author} {\bibfnamefont {J.~M.}\ \bibnamefont
  {{Casandjian}}}, \bibinfo {author} {\bibfnamefont {E.}~\bibnamefont
  {{Charles}}}, \bibinfo {author} {\bibfnamefont {S.}~\bibnamefont {{Chaty}}},
  \bibinfo {author} {\bibfnamefont {A.}~\bibnamefont {{Chekhtman}}}, \bibinfo
  {author} {\bibfnamefont {C.~C.}\ \bibnamefont {{Cheung}}}, \bibinfo {author}
  {\bibfnamefont {J.}~\bibnamefont {{Chiang}}}, \bibinfo {author}
  {\bibfnamefont {S.}~\bibnamefont {{Ciprini}}}, \bibinfo {author}
  {\bibfnamefont {R.}~\bibnamefont {{Claus}}}, \bibinfo {author} {\bibfnamefont
  {J.}~\bibnamefont {{Cohen-Tanugi}}}, \bibinfo {author} {\bibfnamefont
  {J.}~\bibnamefont {{Conrad}}}, \bibinfo {author} {\bibfnamefont {M.~E.}\
  \bibnamefont {{Decesar}}}, \bibinfo {author} {\bibfnamefont {C.~D.}\
  \bibnamefont {{Dermer}}}, \bibinfo {author} {\bibfnamefont {F.}~\bibnamefont
  {{de Palma}}}, \bibinfo {author} {\bibfnamefont {S.~W.}\ \bibnamefont
  {{Digel}}}, \bibinfo {author} {\bibfnamefont {E.~D. C.~E.}\ \bibnamefont
  {{Silva}}}, \bibinfo {author} {\bibfnamefont {P.~S.}\ \bibnamefont
  {{Drell}}}, \bibinfo {author} {\bibfnamefont {R.}~\bibnamefont {{Dubois}}},
  \bibinfo {author} {\bibfnamefont {D.}~\bibnamefont {{Dumora}}}, \bibinfo
  {author} {\bibfnamefont {C.}~\bibnamefont {{Favuzzi}}}, \bibinfo {author}
  {\bibfnamefont {P.}~\bibnamefont {{Fortin}}}, \bibinfo {author}
  {\bibfnamefont {M.}~\bibnamefont {{Frailis}}}, \bibinfo {author}
  {\bibfnamefont {Y.}~\bibnamefont {{Fukazawa}}}, \bibinfo {author}
  {\bibfnamefont {P.}~\bibnamefont {{Fusco}}}, \bibinfo {author} {\bibfnamefont
  {F.}~\bibnamefont {{Gargano}}}, \bibinfo {author} {\bibfnamefont
  {D.}~\bibnamefont {{Gasparrini}}}, \bibinfo {author} {\bibfnamefont
  {N.}~\bibnamefont {{Gehrels}}}, \bibinfo {author} {\bibfnamefont
  {S.}~\bibnamefont {{Germani}}}, \bibinfo {author} {\bibfnamefont
  {N.}~\bibnamefont {{Giglietto}}}, \bibinfo {author} {\bibfnamefont
  {F.}~\bibnamefont {{Giordano}}}, \bibinfo {author} {\bibfnamefont
  {T.}~\bibnamefont {{Glanzman}}}, \bibinfo {author} {\bibfnamefont
  {G.}~\bibnamefont {{Godfrey}}}, \bibinfo {author} {\bibfnamefont
  {I.}~\bibnamefont {{Grenier}}}, \bibinfo {author} {\bibfnamefont {M.~H.}\
  \bibnamefont {{Grondin}}}, \bibinfo {author} {\bibfnamefont {J.~E.}\
  \bibnamefont {{Grove}}}, \bibinfo {author} {\bibfnamefont {L.}~\bibnamefont
  {{Guillemot}}}, \bibinfo {author} {\bibfnamefont {S.}~\bibnamefont
  {{Guiriec}}}, \bibinfo {author} {\bibfnamefont {D.}~\bibnamefont
  {{Hadasch}}}, \bibinfo {author} {\bibfnamefont {A.~K.}\ \bibnamefont
  {{Harding}}}, \bibinfo {author} {\bibfnamefont {E.}~\bibnamefont {{Hays}}},
  \bibinfo {author} {\bibfnamefont {P.}~\bibnamefont {{Jean}}}, \bibinfo
  {author} {\bibfnamefont {G.}~\bibnamefont {{J{\'o}hannesson}}}, \bibinfo
  {author} {\bibfnamefont {T.~J.}\ \bibnamefont {{Johnson}}}, \bibinfo {author}
  {\bibfnamefont {W.~N.}\ \bibnamefont {{Johnson}}}, \bibinfo {author}
  {\bibfnamefont {T.}~\bibnamefont {{Kamae}}}, \bibinfo {author} {\bibfnamefont
  {H.}~\bibnamefont {{Katagiri}}}, \bibinfo {author} {\bibfnamefont
  {J.}~\bibnamefont {{Kataoka}}}, \bibinfo {author} {\bibfnamefont
  {M.}~\bibnamefont {{Kerr}}}, \bibinfo {author} {\bibfnamefont
  {J.}~\bibnamefont {{Kn{\"o}dlseder}}}, \bibinfo {author} {\bibfnamefont
  {M.}~\bibnamefont {{Kuss}}}, \bibinfo {author} {\bibfnamefont
  {J.}~\bibnamefont {{Lande}}}, \bibinfo {author} {\bibfnamefont
  {L.}~\bibnamefont {{Latronico}}}, \bibinfo {author} {\bibfnamefont {S.~H.}\
  \bibnamefont {{Lee}}}, \bibinfo {author} {\bibfnamefont {M.}~\bibnamefont
  {{Lemoine-Goumard}}}, \bibinfo {author} {\bibfnamefont {M.}~\bibnamefont
  {{Llena Garde}}}, \bibinfo {author} {\bibfnamefont {F.}~\bibnamefont
  {{Longo}}}, \bibinfo {author} {\bibfnamefont {F.}~\bibnamefont {{Loparco}}},
  \bibinfo {author} {\bibfnamefont {M.~N.}\ \bibnamefont {{Lovellette}}},
  \bibinfo {author} {\bibfnamefont {P.}~\bibnamefont {{Lubrano}}}, \bibinfo
  {author} {\bibfnamefont {A.}~\bibnamefont {{Makeev}}}, \bibinfo {author}
  {\bibfnamefont {M.~N.}\ \bibnamefont {{Mazziotta}}}, \bibinfo {author}
  {\bibfnamefont {P.~F.}\ \bibnamefont {{Michelson}}}, \bibinfo {author}
  {\bibfnamefont {W.}~\bibnamefont {{Mitthumsiri}}}, \bibinfo {author}
  {\bibfnamefont {T.}~\bibnamefont {{Mizuno}}}, \bibinfo {author}
  {\bibfnamefont {C.}~\bibnamefont {{Monte}}}, \bibinfo {author} {\bibfnamefont
  {M.~E.}\ \bibnamefont {{Monzani}}}, \bibinfo {author} {\bibfnamefont
  {A.}~\bibnamefont {{Morselli}}}, \bibinfo {author} {\bibfnamefont {I.~V.}\
  \bibnamefont {{Moskalenko}}}, \bibinfo {author} {\bibfnamefont
  {S.}~\bibnamefont {{Murgia}}}, \bibinfo {author} {\bibfnamefont
  {M.}~\bibnamefont {{Naumann-Godo}}}, \bibinfo {author} {\bibfnamefont
  {P.~L.}\ \bibnamefont {{Nolan}}}, \bibinfo {author} {\bibfnamefont {J.~P.}\
  \bibnamefont {{Norris}}}, \bibinfo {author} {\bibfnamefont {E.}~\bibnamefont
  {{Nuss}}}, \bibinfo {author} {\bibfnamefont {T.}~\bibnamefont {{Ohsugi}}},
  \bibinfo {author} {\bibfnamefont {N.}~\bibnamefont {{Omodei}}}, \bibinfo
  {author} {\bibfnamefont {E.}~\bibnamefont {{Orlando}}}, \bibinfo {author}
  {\bibfnamefont {J.~F.}\ \bibnamefont {{Ormes}}}, \bibinfo {author}
  {\bibfnamefont {B.}~\bibnamefont {{Pancrazi}}}, \bibinfo {author}
  {\bibfnamefont {D.}~\bibnamefont {{Parent}}}, \bibinfo {author}
  {\bibfnamefont {M.}~\bibnamefont {{Pepe}}}, \bibinfo {author} {\bibfnamefont
  {M.}~\bibnamefont {{Pesce-Rollins}}}, \bibinfo {author} {\bibfnamefont
  {F.}~\bibnamefont {{Piron}}}, \bibinfo {author} {\bibfnamefont {T.~A.}\
  \bibnamefont {{Porter}}}, \bibinfo {author} {\bibfnamefont {S.}~\bibnamefont
  {{Rain{\`o}}}}, \bibinfo {author} {\bibfnamefont {R.}~\bibnamefont
  {{Rando}}}, \bibinfo {author} {\bibfnamefont {A.}~\bibnamefont {{Reimer}}},
  \bibinfo {author} {\bibfnamefont {O.}~\bibnamefont {{Reimer}}}, \bibinfo
  {author} {\bibfnamefont {T.}~\bibnamefont {{Reposeur}}}, \bibinfo {author}
  {\bibfnamefont {J.}~\bibnamefont {{Ripken}}}, \bibinfo {author}
  {\bibfnamefont {R.~W.}\ \bibnamefont {{Romani}}}, \bibinfo {author}
  {\bibfnamefont {M.}~\bibnamefont {{Roth}}}, \bibinfo {author} {\bibfnamefont
  {H.~F.~W.}\ \bibnamefont {{Sadrozinski}}}, \bibinfo {author} {\bibfnamefont
  {P.~M.}\ \bibnamefont {{Saz Parkinson}}}, \bibinfo {author} {\bibfnamefont
  {C.}~\bibnamefont {{Sgr{\`o}}}}, \bibinfo {author} {\bibfnamefont {E.~J.}\
  \bibnamefont {{Siskind}}}, \bibinfo {author} {\bibfnamefont {D.~A.}\
  \bibnamefont {{Smith}}}, \bibinfo {author} {\bibfnamefont {P.}~\bibnamefont
  {{Spinelli}}}, \bibinfo {author} {\bibfnamefont {M.~S.}\ \bibnamefont
  {{Strickman}}}, \bibinfo {author} {\bibfnamefont {D.~J.}\ \bibnamefont
  {{Suson}}}, \bibinfo {author} {\bibfnamefont {H.}~\bibnamefont
  {{Takahashi}}}, \bibinfo {author} {\bibfnamefont {T.}~\bibnamefont
  {{Takahashi}}}, \bibinfo {author} {\bibfnamefont {T.}~\bibnamefont
  {{Tanaka}}}, \bibinfo {author} {\bibfnamefont {J.~B.}\ \bibnamefont
  {{Thayer}}}, \bibinfo {author} {\bibfnamefont {J.~G.}\ \bibnamefont
  {{Thayer}}}, \bibinfo {author} {\bibfnamefont {L.}~\bibnamefont {{Tibaldo}}},
  \bibinfo {author} {\bibfnamefont {D.~F.}\ \bibnamefont {{Torres}}}, \bibinfo
  {author} {\bibfnamefont {G.}~\bibnamefont {{Tosti}}}, \bibinfo {author}
  {\bibfnamefont {A.}~\bibnamefont {{Tramacere}}}, \bibinfo {author}
  {\bibfnamefont {Y.}~\bibnamefont {{Uchiyama}}}, \bibinfo {author}
  {\bibfnamefont {T.~L.}\ \bibnamefont {{Usher}}}, \bibinfo {author}
  {\bibfnamefont {V.}~\bibnamefont {{Vasileiou}}}, \bibinfo {author}
  {\bibfnamefont {C.}~\bibnamefont {{Venter}}}, \bibinfo {author}
  {\bibfnamefont {N.}~\bibnamefont {{Vilchez}}}, \bibinfo {author}
  {\bibfnamefont {V.}~\bibnamefont {{Vitale}}}, \bibinfo {author}
  {\bibfnamefont {A.~P.}\ \bibnamefont {{Waite}}}, \bibinfo {author}
  {\bibfnamefont {P.}~\bibnamefont {{Wang}}}, \bibinfo {author} {\bibfnamefont
  {N.}~\bibnamefont {{Webb}}}, \bibinfo {author} {\bibfnamefont {B.~L.}\
  \bibnamefont {{Winer}}}, \bibinfo {author} {\bibfnamefont {Z.}~\bibnamefont
  {{Yang}}}, \bibinfo {author} {\bibfnamefont {T.}~\bibnamefont {{Ylinen}}},
  \bibinfo {author} {\bibfnamefont {M.}~\bibnamefont {{Ziegler}}},\ and\
  \bibinfo {author} {\bibnamefont {{Fermi LAT Collaboration}}},\ }\bibfield
  {title} {\bibinfo {title} {{A population of gamma-ray emitting globular
  clusters seen with the Fermi Large Area Telescope}},\ }\href
  {https://doi.org/10.1051/0004-6361/201014458} {\bibfield  {journal} {\bibinfo
   {journal} {Astronomy and Astrophysics}\ }\textbf {\bibinfo {volume} {524}},\
  \bibinfo {eid} {A75} (\bibinfo {year} {2010})},\ \Eprint
  {https://arxiv.org/abs/1003.3588} {arXiv:1003.3588 [astro-ph.GA]}
  \BibitemShut {NoStop}%
\bibitem [{\citenamefont {Conrad}\ \emph {et~al.}(2015)\citenamefont {Conrad},
  \citenamefont {Cohen-Tanugi},\ and\ \citenamefont
  {Strigari}}]{Conrad:2015bsa}%
  \BibitemOpen
  \bibfield  {author} {\bibinfo {author} {\bibfnamefont {J.}~\bibnamefont
  {Conrad}}, \bibinfo {author} {\bibfnamefont {J.}~\bibnamefont
  {Cohen-Tanugi}},\ and\ \bibinfo {author} {\bibfnamefont {L.~E.}\ \bibnamefont
  {Strigari}},\ }\bibfield  {title} {\bibinfo {title} {{WIMP searches with
  gamma rays in the Fermi era: challenges, methods and results}},\ }\href
  {https://doi.org/10.1134/S1063776115130099} {\bibfield  {journal} {\bibinfo
  {journal} {J. Exp. Theor. Phys.}\ }\textbf {\bibinfo {volume} {121}},\
  \bibinfo {pages} {1104} (\bibinfo {year} {2015})},\ \Eprint
  {https://arxiv.org/abs/1503.06348} {arXiv:1503.06348 [astro-ph.CO]}
  \BibitemShut {NoStop}%
\bibitem [{\citenamefont {Strigari}(2018)}]{Strigari:2018utn}%
  \BibitemOpen
  \bibfield  {author} {\bibinfo {author} {\bibfnamefont {L.~E.}\ \bibnamefont
  {Strigari}},\ }\bibfield  {title} {\bibinfo {title} {{Dark matter in dwarf
  spheroidal galaxies and indirect detection: a review}},\ }\href
  {https://doi.org/10.1088/1361-6633/aaae16} {\bibfield  {journal} {\bibinfo
  {journal} {Rept. Prog. Phys.}\ }\textbf {\bibinfo {volume} {81}},\ \bibinfo
  {pages} {056901} (\bibinfo {year} {2018})},\ \Eprint
  {https://arxiv.org/abs/1805.05883} {arXiv:1805.05883 [astro-ph.CO]}
  \BibitemShut {NoStop}%
\bibitem [{\citenamefont {{Drlica-Wagner}}\ \emph {et~al.}(2019)\citenamefont
  {{Drlica-Wagner}}, \citenamefont {{Mao}}, \citenamefont {{Adhikari}},
  \citenamefont {{Armstrong}}, \citenamefont {{Banerjee}}, \citenamefont
  {{Banik}}, \citenamefont {{Bechtol}}, \citenamefont {{Bird}}, \citenamefont
  {{Boddy}}, \citenamefont {{Bonaca}}, \citenamefont {{Bovy}}, \citenamefont
  {{Buckley}}, \citenamefont {{Bulbul}}, \citenamefont {{Chang}}, \citenamefont
  {{Chapline}}, \citenamefont {{Cohen-Tanugi}}, \citenamefont {{Cuoco}},
  \citenamefont {{Cyr-Racine}}, \citenamefont {{Dawson}}, \citenamefont
  {{D{\'\i}az Rivero}}, \citenamefont {{Dvorkin}}, \citenamefont {{Erkal}},
  \citenamefont {{Fassnacht}}, \citenamefont {{Garc{\'\i}a-Bellido}},
  \citenamefont {{Giannotti}}, \citenamefont {{Gluscevic}}, \citenamefont
  {{Golovich}}, \citenamefont {{Hendel}}, \citenamefont {{Hezaveh}},
  \citenamefont {{Horiuchi}}, \citenamefont {{Jee}}, \citenamefont
  {{Kaplinghat}}, \citenamefont {{Keeton}}, \citenamefont {{Koposov}},
  \citenamefont {{Lam}}, \citenamefont {{Li}}, \citenamefont {{Lu}},
  \citenamefont {{Mandelbaum}}, \citenamefont {{McDermott}}, \citenamefont
  {{McNanna}}, \citenamefont {{Medford}}, \citenamefont {{Meyer}},
  \citenamefont {{Marc}}, \citenamefont {{Murgia}}, \citenamefont {{Nadler}},
  \citenamefont {{Necib}}, \citenamefont {{Nuss}}, \citenamefont {{Pace}},
  \citenamefont {{Peter}}, \citenamefont {{Polin}}, \citenamefont
  {{Prescod-Weinstein}}, \citenamefont {{Read}}, \citenamefont {{Rosenfeld}},
  \citenamefont {{Shipp}}, \citenamefont {{Simon}}, \citenamefont {{Slatyer}},
  \citenamefont {{Straniero}}, \citenamefont {{Strigari}}, \citenamefont
  {{Tollerud}}, \citenamefont {{Tyson}}, \citenamefont {{Wang}}, \citenamefont
  {{Wechsler}}, \citenamefont {{Wittman}}, \citenamefont {{Yu}}, \citenamefont
  {{Zaharijas}}, \citenamefont {{Ali-Ha{\"\i}moud}}, \citenamefont {{Annis}},
  \citenamefont {{Birrer}}, \citenamefont {{Biswas}}, \citenamefont {{Blazek}},
  \citenamefont {{Brooks}}, \citenamefont {{Buckley-Geer}}, \citenamefont
  {{Caputo}}, \citenamefont {{Charles}}, \citenamefont {{Digel}}, \citenamefont
  {{Dodelson}}, \citenamefont {{Flaugher}}, \citenamefont {{Frieman}},
  \citenamefont {{Gawiser}}, \citenamefont {{Hearin}}, \citenamefont
  {{Hlo{\v{z}}ek}}, \citenamefont {{Jain}}, \citenamefont {{Jeltema}},
  \citenamefont {{Koushiappas}}, \citenamefont {{Lisanti}}, \citenamefont
  {{LoVerde}}, \citenamefont {{Mishra-Sharma}}, \citenamefont {{Newman}},
  \citenamefont {{Nord}}, \citenamefont {{Nourbakhsh}}, \citenamefont {{Ritz}},
  \citenamefont {{Robertson}}, \citenamefont {{S{\'a}nchez-Conde}},
  \citenamefont {{Slosar}}, \citenamefont {{Tait}}, \citenamefont {{Verma}},
  \citenamefont {{Vilalta}}, \citenamefont {{Walter}}, \citenamefont
  {{Yanny}},\ and\ \citenamefont {{Zentner}}}]{2019arXiv190201055D}%
  \BibitemOpen
  \bibfield  {author} {\bibinfo {author} {\bibfnamefont {A.}~\bibnamefont
  {{Drlica-Wagner}}}, \bibinfo {author} {\bibfnamefont {Y.-Y.}\ \bibnamefont
  {{Mao}}}, \bibinfo {author} {\bibfnamefont {S.}~\bibnamefont {{Adhikari}}},
  \bibinfo {author} {\bibfnamefont {R.}~\bibnamefont {{Armstrong}}}, \bibinfo
  {author} {\bibfnamefont {A.}~\bibnamefont {{Banerjee}}}, \bibinfo {author}
  {\bibfnamefont {N.}~\bibnamefont {{Banik}}}, \bibinfo {author} {\bibfnamefont
  {K.}~\bibnamefont {{Bechtol}}}, \bibinfo {author} {\bibfnamefont
  {S.}~\bibnamefont {{Bird}}}, \bibinfo {author} {\bibfnamefont {K.~K.}\
  \bibnamefont {{Boddy}}}, \bibinfo {author} {\bibfnamefont {A.}~\bibnamefont
  {{Bonaca}}}, \bibinfo {author} {\bibfnamefont {J.}~\bibnamefont {{Bovy}}},
  \bibinfo {author} {\bibfnamefont {M.~R.}\ \bibnamefont {{Buckley}}}, \bibinfo
  {author} {\bibfnamefont {E.}~\bibnamefont {{Bulbul}}}, \bibinfo {author}
  {\bibfnamefont {C.}~\bibnamefont {{Chang}}}, \bibinfo {author} {\bibfnamefont
  {G.}~\bibnamefont {{Chapline}}}, \bibinfo {author} {\bibfnamefont
  {J.}~\bibnamefont {{Cohen-Tanugi}}}, \bibinfo {author} {\bibfnamefont
  {A.}~\bibnamefont {{Cuoco}}}, \bibinfo {author} {\bibfnamefont {F.-Y.}\
  \bibnamefont {{Cyr-Racine}}}, \bibinfo {author} {\bibfnamefont {W.~A.}\
  \bibnamefont {{Dawson}}}, \bibinfo {author} {\bibfnamefont {A.}~\bibnamefont
  {{D{\'\i}az Rivero}}}, \bibinfo {author} {\bibfnamefont {C.}~\bibnamefont
  {{Dvorkin}}}, \bibinfo {author} {\bibfnamefont {D.}~\bibnamefont {{Erkal}}},
  \bibinfo {author} {\bibfnamefont {C.~D.}\ \bibnamefont {{Fassnacht}}},
  \bibinfo {author} {\bibfnamefont {J.}~\bibnamefont {{Garc{\'\i}a-Bellido}}},
  \bibinfo {author} {\bibfnamefont {M.}~\bibnamefont {{Giannotti}}}, \bibinfo
  {author} {\bibfnamefont {V.}~\bibnamefont {{Gluscevic}}}, \bibinfo {author}
  {\bibfnamefont {N.}~\bibnamefont {{Golovich}}}, \bibinfo {author}
  {\bibfnamefont {D.}~\bibnamefont {{Hendel}}}, \bibinfo {author}
  {\bibfnamefont {Y.~D.}\ \bibnamefont {{Hezaveh}}}, \bibinfo {author}
  {\bibfnamefont {S.}~\bibnamefont {{Horiuchi}}}, \bibinfo {author}
  {\bibfnamefont {M.~J.}\ \bibnamefont {{Jee}}}, \bibinfo {author}
  {\bibfnamefont {M.}~\bibnamefont {{Kaplinghat}}}, \bibinfo {author}
  {\bibfnamefont {C.~R.}\ \bibnamefont {{Keeton}}}, \bibinfo {author}
  {\bibfnamefont {S.~E.}\ \bibnamefont {{Koposov}}}, \bibinfo {author}
  {\bibfnamefont {C.~Y.}\ \bibnamefont {{Lam}}}, \bibinfo {author}
  {\bibfnamefont {T.~S.}\ \bibnamefont {{Li}}}, \bibinfo {author}
  {\bibfnamefont {J.~R.}\ \bibnamefont {{Lu}}}, \bibinfo {author}
  {\bibfnamefont {R.}~\bibnamefont {{Mandelbaum}}}, \bibinfo {author}
  {\bibfnamefont {S.~D.}\ \bibnamefont {{McDermott}}}, \bibinfo {author}
  {\bibfnamefont {M.}~\bibnamefont {{McNanna}}}, \bibinfo {author}
  {\bibfnamefont {M.}~\bibnamefont {{Medford}}}, \bibinfo {author}
  {\bibfnamefont {M.}~\bibnamefont {{Meyer}}}, \bibinfo {author} {\bibfnamefont
  {M.}~\bibnamefont {{Marc}}}, \bibinfo {author} {\bibfnamefont
  {S.}~\bibnamefont {{Murgia}}}, \bibinfo {author} {\bibfnamefont {E.~O.}\
  \bibnamefont {{Nadler}}}, \bibinfo {author} {\bibfnamefont {L.}~\bibnamefont
  {{Necib}}}, \bibinfo {author} {\bibfnamefont {E.}~\bibnamefont {{Nuss}}},
  \bibinfo {author} {\bibfnamefont {A.~B.}\ \bibnamefont {{Pace}}}, \bibinfo
  {author} {\bibfnamefont {A.~H.~G.}\ \bibnamefont {{Peter}}}, \bibinfo
  {author} {\bibfnamefont {D.~A.}\ \bibnamefont {{Polin}}}, \bibinfo {author}
  {\bibfnamefont {C.}~\bibnamefont {{Prescod-Weinstein}}}, \bibinfo {author}
  {\bibfnamefont {J.~I.}\ \bibnamefont {{Read}}}, \bibinfo {author}
  {\bibfnamefont {R.}~\bibnamefont {{Rosenfeld}}}, \bibinfo {author}
  {\bibfnamefont {N.}~\bibnamefont {{Shipp}}}, \bibinfo {author} {\bibfnamefont
  {J.~D.}\ \bibnamefont {{Simon}}}, \bibinfo {author} {\bibfnamefont {T.~R.}\
  \bibnamefont {{Slatyer}}}, \bibinfo {author} {\bibfnamefont {O.}~\bibnamefont
  {{Straniero}}}, \bibinfo {author} {\bibfnamefont {L.~E.}\ \bibnamefont
  {{Strigari}}}, \bibinfo {author} {\bibfnamefont {E.}~\bibnamefont
  {{Tollerud}}}, \bibinfo {author} {\bibfnamefont {J.~A.}\ \bibnamefont
  {{Tyson}}}, \bibinfo {author} {\bibfnamefont {M.-Y.}\ \bibnamefont {{Wang}}},
  \bibinfo {author} {\bibfnamefont {R.~H.}\ \bibnamefont {{Wechsler}}},
  \bibinfo {author} {\bibfnamefont {D.}~\bibnamefont {{Wittman}}}, \bibinfo
  {author} {\bibfnamefont {H.-B.}\ \bibnamefont {{Yu}}}, \bibinfo {author}
  {\bibfnamefont {G.}~\bibnamefont {{Zaharijas}}}, \bibinfo {author}
  {\bibfnamefont {Y.}~\bibnamefont {{Ali-Ha{\"\i}moud}}}, \bibinfo {author}
  {\bibfnamefont {J.}~\bibnamefont {{Annis}}}, \bibinfo {author} {\bibfnamefont
  {S.}~\bibnamefont {{Birrer}}}, \bibinfo {author} {\bibfnamefont
  {R.}~\bibnamefont {{Biswas}}}, \bibinfo {author} {\bibfnamefont
  {J.}~\bibnamefont {{Blazek}}}, \bibinfo {author} {\bibfnamefont {A.~M.}\
  \bibnamefont {{Brooks}}}, \bibinfo {author} {\bibfnamefont {E.}~\bibnamefont
  {{Buckley-Geer}}}, \bibinfo {author} {\bibfnamefont {R.}~\bibnamefont
  {{Caputo}}}, \bibinfo {author} {\bibfnamefont {E.}~\bibnamefont {{Charles}}},
  \bibinfo {author} {\bibfnamefont {S.}~\bibnamefont {{Digel}}}, \bibinfo
  {author} {\bibfnamefont {S.}~\bibnamefont {{Dodelson}}}, \bibinfo {author}
  {\bibfnamefont {B.}~\bibnamefont {{Flaugher}}}, \bibinfo {author}
  {\bibfnamefont {J.}~\bibnamefont {{Frieman}}}, \bibinfo {author}
  {\bibfnamefont {E.}~\bibnamefont {{Gawiser}}}, \bibinfo {author}
  {\bibfnamefont {A.~P.}\ \bibnamefont {{Hearin}}}, \bibinfo {author}
  {\bibfnamefont {R.}~\bibnamefont {{Hlo{\v{z}}ek}}}, \bibinfo {author}
  {\bibfnamefont {B.}~\bibnamefont {{Jain}}}, \bibinfo {author} {\bibfnamefont
  {T.~E.}\ \bibnamefont {{Jeltema}}}, \bibinfo {author} {\bibfnamefont {S.~M.}\
  \bibnamefont {{Koushiappas}}}, \bibinfo {author} {\bibfnamefont
  {M.}~\bibnamefont {{Lisanti}}}, \bibinfo {author} {\bibfnamefont
  {M.}~\bibnamefont {{LoVerde}}}, \bibinfo {author} {\bibfnamefont
  {S.}~\bibnamefont {{Mishra-Sharma}}}, \bibinfo {author} {\bibfnamefont
  {J.~A.}\ \bibnamefont {{Newman}}}, \bibinfo {author} {\bibfnamefont
  {B.}~\bibnamefont {{Nord}}}, \bibinfo {author} {\bibfnamefont
  {E.}~\bibnamefont {{Nourbakhsh}}}, \bibinfo {author} {\bibfnamefont
  {S.}~\bibnamefont {{Ritz}}}, \bibinfo {author} {\bibfnamefont {B.~E.}\
  \bibnamefont {{Robertson}}}, \bibinfo {author} {\bibfnamefont {M.~A.}\
  \bibnamefont {{S{\'a}nchez-Conde}}}, \bibinfo {author} {\bibfnamefont
  {A.}~\bibnamefont {{Slosar}}}, \bibinfo {author} {\bibfnamefont {T.~M.~P.}\
  \bibnamefont {{Tait}}}, \bibinfo {author} {\bibfnamefont {A.}~\bibnamefont
  {{Verma}}}, \bibinfo {author} {\bibfnamefont {R.}~\bibnamefont {{Vilalta}}},
  \bibinfo {author} {\bibfnamefont {C.~W.}\ \bibnamefont {{Walter}}}, \bibinfo
  {author} {\bibfnamefont {B.}~\bibnamefont {{Yanny}}},\ and\ \bibinfo {author}
  {\bibfnamefont {A.~R.}\ \bibnamefont {{Zentner}}},\ }\bibfield  {title}
  {\bibinfo {title} {{Probing the Fundamental Nature of Dark Matter with the
  Large Synoptic Survey Telescope}},\ }\href
  {https://doi.org/10.48550/arXiv.1902.01055} {\bibfield  {journal} {\bibinfo
  {journal} {arXiv e-prints}\ ,\ \bibinfo {eid} {arXiv:1902.01055}} (\bibinfo
  {year} {2019})},\ \Eprint {https://arxiv.org/abs/1902.01055}
  {arXiv:1902.01055 [astro-ph.CO]} \BibitemShut {NoStop}%
\bibitem [{\citenamefont {{Simon}}\ and\ \citenamefont
  {{Geha}}(2007)}]{Simon:2007dq}%
  \BibitemOpen
  \bibfield  {author} {\bibinfo {author} {\bibfnamefont {J.~D.}\ \bibnamefont
  {{Simon}}}\ and\ \bibinfo {author} {\bibfnamefont {M.}~\bibnamefont
  {{Geha}}},\ }\bibfield  {title} {\bibinfo {title} {{The Kinematics of the
  Ultra-faint Milky Way Satellites: Solving the Missing Satellite Problem}},\
  }\href {https://doi.org/10.1086/521816} {\bibfield  {journal} {\bibinfo
  {journal} {\apj}\ }\textbf {\bibinfo {volume} {670}},\ \bibinfo {pages} {313}
  (\bibinfo {year} {2007})},\ \Eprint {https://arxiv.org/abs/0706.0516}
  {arXiv:0706.0516 [astro-ph]} \BibitemShut {NoStop}%
\bibitem [{\citenamefont {Simon}(2019)}]{Simon:2019nxf}%
  \BibitemOpen
  \bibfield  {author} {\bibinfo {author} {\bibfnamefont {J.~D.}\ \bibnamefont
  {Simon}},\ }\bibfield  {title} {\bibinfo {title} {{The Faintest Dwarf
  Galaxies}},\ }\href {https://doi.org/10.1146/annurev-astro-091918-104453}
  {\bibfield  {journal} {\bibinfo  {journal} {Ann. Rev. Astron. Astrophys.}\
  }\textbf {\bibinfo {volume} {57}},\ \bibinfo {pages} {375} (\bibinfo {year}
  {2019})},\ \Eprint {https://arxiv.org/abs/1901.05465} {arXiv:1901.05465
  [astro-ph.GA]} \BibitemShut {NoStop}%
\bibitem [{\citenamefont {Strigari}\ \emph {et~al.}(2008)\citenamefont
  {Strigari}, \citenamefont {Koushiappas}, \citenamefont {Bullock},
  \citenamefont {Kaplinghat}, \citenamefont {Simon}, \citenamefont {Geha},\
  and\ \citenamefont {Willman}}]{Strigari:2007at}%
  \BibitemOpen
  \bibfield  {author} {\bibinfo {author} {\bibfnamefont {L.~E.}\ \bibnamefont
  {Strigari}}, \bibinfo {author} {\bibfnamefont {S.~M.}\ \bibnamefont
  {Koushiappas}}, \bibinfo {author} {\bibfnamefont {J.~S.}\ \bibnamefont
  {Bullock}}, \bibinfo {author} {\bibfnamefont {M.}~\bibnamefont {Kaplinghat}},
  \bibinfo {author} {\bibfnamefont {J.~D.}\ \bibnamefont {Simon}}, \bibinfo
  {author} {\bibfnamefont {M.}~\bibnamefont {Geha}},\ and\ \bibinfo {author}
  {\bibfnamefont {B.}~\bibnamefont {Willman}},\ }\bibfield  {title} {\bibinfo
  {title} {{The Most Dark Matter Dominated Galaxies: Predicted Gamma-ray
  Signals from the Faintest Milky Way Dwarfs}},\ }\href
  {https://doi.org/10.1086/529488} {\bibfield  {journal} {\bibinfo  {journal}
  {Astrophys. J.}\ }\textbf {\bibinfo {volume} {678}},\ \bibinfo {pages} {614}
  (\bibinfo {year} {2008})},\ \Eprint {https://arxiv.org/abs/0709.1510}
  {arXiv:0709.1510 [astro-ph]} \BibitemShut {NoStop}%
\bibitem [{\citenamefont {Walker}\ \emph {et~al.}(2009)\citenamefont {Walker},
  \citenamefont {Mateo}, \citenamefont {Olszewski}, \citenamefont {Penarrubia},
  \citenamefont {Evans},\ and\ \citenamefont {Gilmore}}]{Walker:2009zp}%
  \BibitemOpen
  \bibfield  {author} {\bibinfo {author} {\bibfnamefont {M.~G.}\ \bibnamefont
  {Walker}}, \bibinfo {author} {\bibfnamefont {M.}~\bibnamefont {Mateo}},
  \bibinfo {author} {\bibfnamefont {E.~W.}\ \bibnamefont {Olszewski}}, \bibinfo
  {author} {\bibfnamefont {J.}~\bibnamefont {Penarrubia}}, \bibinfo {author}
  {\bibfnamefont {N.~W.}\ \bibnamefont {Evans}},\ and\ \bibinfo {author}
  {\bibfnamefont {G.}~\bibnamefont {Gilmore}},\ }\bibfield  {title} {\bibinfo
  {title} {{A Universal Mass Profile for Dwarf Spheroidal Galaxies}},\ }\href
  {https://doi.org/10.1088/0004-637X/704/2/1274} {\bibfield  {journal}
  {\bibinfo  {journal} {Astrophys. J.}\ }\textbf {\bibinfo {volume} {704}},\
  \bibinfo {pages} {1274} (\bibinfo {year} {2009})},\ \bibinfo {note}
  {[Erratum: Astrophys.J. 710, 886--890 (2010)]},\ \Eprint
  {https://arxiv.org/abs/0906.0341} {arXiv:0906.0341 [astro-ph.CO]}
  \BibitemShut {NoStop}%
\bibitem [{\citenamefont {Navarro}\ \emph {et~al.}(1997)\citenamefont
  {Navarro}, \citenamefont {Frenk},\ and\ \citenamefont
  {White}}]{Navarro:1996gj}%
  \BibitemOpen
  \bibfield  {author} {\bibinfo {author} {\bibfnamefont {J.~F.}\ \bibnamefont
  {Navarro}}, \bibinfo {author} {\bibfnamefont {C.~S.}\ \bibnamefont {Frenk}},\
  and\ \bibinfo {author} {\bibfnamefont {S.~D.~M.}\ \bibnamefont {White}},\
  }\bibfield  {title} {\bibinfo {title} {{A Universal density profile from
  hierarchical clustering}},\ }\href {https://doi.org/10.1086/304888}
  {\bibfield  {journal} {\bibinfo  {journal} {Astrophys. J.}\ }\textbf
  {\bibinfo {volume} {490}},\ \bibinfo {pages} {493} (\bibinfo {year}
  {1997})},\ \Eprint {https://arxiv.org/abs/astro-ph/9611107}
  {arXiv:astro-ph/9611107} \BibitemShut {NoStop}%
\bibitem [{\citenamefont {{Smith}}\ \emph {et~al.}(2024)\citenamefont
  {{Smith}}, \citenamefont {{Cerny}}, \citenamefont {{Hayes}}, \citenamefont
  {{Sestito}}, \citenamefont {{Jensen}}, \citenamefont {{McConnachie}},
  \citenamefont {{Geha}}, \citenamefont {{Navarro}}, \citenamefont {{Li}},
  \citenamefont {{Cuillandre}}, \citenamefont {{Errani}}, \citenamefont
  {{Chambers}}, \citenamefont {{Gwyn}}, \citenamefont {{Hammer}}, \citenamefont
  {{Hudson}}, \citenamefont {{Magnier}},\ and\ \citenamefont
  {{Martin}}}]{2024ApJ...961...92S}%
  \BibitemOpen
  \bibfield  {author} {\bibinfo {author} {\bibfnamefont {S.~E.~T.}\
  \bibnamefont {{Smith}}}, \bibinfo {author} {\bibfnamefont {W.}~\bibnamefont
  {{Cerny}}}, \bibinfo {author} {\bibfnamefont {C.~R.}\ \bibnamefont
  {{Hayes}}}, \bibinfo {author} {\bibfnamefont {F.}~\bibnamefont {{Sestito}}},
  \bibinfo {author} {\bibfnamefont {J.}~\bibnamefont {{Jensen}}}, \bibinfo
  {author} {\bibfnamefont {A.~W.}\ \bibnamefont {{McConnachie}}}, \bibinfo
  {author} {\bibfnamefont {M.}~\bibnamefont {{Geha}}}, \bibinfo {author}
  {\bibfnamefont {J.~F.}\ \bibnamefont {{Navarro}}}, \bibinfo {author}
  {\bibfnamefont {T.~S.}\ \bibnamefont {{Li}}}, \bibinfo {author}
  {\bibfnamefont {J.-C.}\ \bibnamefont {{Cuillandre}}}, \bibinfo {author}
  {\bibfnamefont {R.}~\bibnamefont {{Errani}}}, \bibinfo {author}
  {\bibfnamefont {K.}~\bibnamefont {{Chambers}}}, \bibinfo {author}
  {\bibfnamefont {S.}~\bibnamefont {{Gwyn}}}, \bibinfo {author} {\bibfnamefont
  {F.}~\bibnamefont {{Hammer}}}, \bibinfo {author} {\bibfnamefont {M.~J.}\
  \bibnamefont {{Hudson}}}, \bibinfo {author} {\bibfnamefont {E.}~\bibnamefont
  {{Magnier}}},\ and\ \bibinfo {author} {\bibfnamefont {N.}~\bibnamefont
  {{Martin}}},\ }\bibfield  {title} {\bibinfo {title} {{The Discovery of the
  Faintest Known Milky Way Satellite Using UNIONS}},\ }\href
  {https://doi.org/10.3847/1538-4357/ad0d9f} {\bibfield  {journal} {\bibinfo
  {journal} {\apj}\ }\textbf {\bibinfo {volume} {961}},\ \bibinfo {eid} {92}
  (\bibinfo {year} {2024})},\ \Eprint {https://arxiv.org/abs/2311.10147}
  {arXiv:2311.10147 [astro-ph.GA]} \BibitemShut {NoStop}%
\bibitem [{\citenamefont {{Errani}}\ \emph {et~al.}(2024)\citenamefont
  {{Errani}}, \citenamefont {{Navarro}}, \citenamefont {{Smith}},\ and\
  \citenamefont {{McConnachie}}}]{2024ApJ...965...20E}%
  \BibitemOpen
  \bibfield  {author} {\bibinfo {author} {\bibfnamefont {R.}~\bibnamefont
  {{Errani}}}, \bibinfo {author} {\bibfnamefont {J.~F.}\ \bibnamefont
  {{Navarro}}}, \bibinfo {author} {\bibfnamefont {S.~E.~T.}\ \bibnamefont
  {{Smith}}},\ and\ \bibinfo {author} {\bibfnamefont {A.~W.}\ \bibnamefont
  {{McConnachie}}},\ }\bibfield  {title} {\bibinfo {title} {{Ursa Major
  III/UNIONS 1: The Darkest Galaxy Ever Discovered?}},\ }\href
  {https://doi.org/10.3847/1538-4357/ad2267} {\bibfield  {journal} {\bibinfo
  {journal} {\apj}\ }\textbf {\bibinfo {volume} {965}},\ \bibinfo {eid} {20}
  (\bibinfo {year} {2024})},\ \Eprint {https://arxiv.org/abs/2311.10134}
  {arXiv:2311.10134 [astro-ph.GA]} \BibitemShut {NoStop}%
\bibitem [{\citenamefont {Goldstein}\ \emph {et~al.}(2022)\citenamefont
  {Goldstein}, \citenamefont {Koushiappas},\ and\ \citenamefont
  {Walker}}]{Goldstein:2022pxu}%
  \BibitemOpen
  \bibfield  {author} {\bibinfo {author} {\bibfnamefont {I.~S.}\ \bibnamefont
  {Goldstein}}, \bibinfo {author} {\bibfnamefont {S.~M.}\ \bibnamefont
  {Koushiappas}},\ and\ \bibinfo {author} {\bibfnamefont {M.~G.}\ \bibnamefont
  {Walker}},\ }\bibfield  {title} {\bibinfo {title} {{Viability of ultralight
  bosonic dark matter in dwarf galaxies}},\ }\href
  {https://doi.org/10.1103/PhysRevD.106.063010} {\bibfield  {journal} {\bibinfo
   {journal} {Phys. Rev. D}\ }\textbf {\bibinfo {volume} {106}},\ \bibinfo
  {pages} {063010} (\bibinfo {year} {2022})},\ \Eprint
  {https://arxiv.org/abs/2206.05244} {arXiv:2206.05244 [astro-ph.GA]}
  \BibitemShut {NoStop}%
\bibitem [{\citenamefont {{Adhikari}}\ \emph {et~al.}(2022)\citenamefont
  {{Adhikari}}, \citenamefont {{Banerjee}}, \citenamefont {{Boddy}},
  \citenamefont {{Cyr-Racine}}, \citenamefont {{Desmond}}, \citenamefont
  {{Dvorkin}}, \citenamefont {{Jain}}, \citenamefont {{Kahlhoefer}},
  \citenamefont {{Kaplinghat}}, \citenamefont {{Nierenberg}}, \citenamefont
  {{Peter}}, \citenamefont {{Robertson}}, \citenamefont {{Sakstein}},\ and\
  \citenamefont {{Zavala}}}]{2022arXiv220710638A}%
  \BibitemOpen
  \bibfield  {author} {\bibinfo {author} {\bibfnamefont {S.}~\bibnamefont
  {{Adhikari}}}, \bibinfo {author} {\bibfnamefont {A.}~\bibnamefont
  {{Banerjee}}}, \bibinfo {author} {\bibfnamefont {K.~K.}\ \bibnamefont
  {{Boddy}}}, \bibinfo {author} {\bibfnamefont {F.-Y.}\ \bibnamefont
  {{Cyr-Racine}}}, \bibinfo {author} {\bibfnamefont {H.}~\bibnamefont
  {{Desmond}}}, \bibinfo {author} {\bibfnamefont {C.}~\bibnamefont
  {{Dvorkin}}}, \bibinfo {author} {\bibfnamefont {B.}~\bibnamefont {{Jain}}},
  \bibinfo {author} {\bibfnamefont {F.}~\bibnamefont {{Kahlhoefer}}}, \bibinfo
  {author} {\bibfnamefont {M.}~\bibnamefont {{Kaplinghat}}}, \bibinfo {author}
  {\bibfnamefont {A.}~\bibnamefont {{Nierenberg}}}, \bibinfo {author}
  {\bibfnamefont {A.~H.~G.}\ \bibnamefont {{Peter}}}, \bibinfo {author}
  {\bibfnamefont {A.}~\bibnamefont {{Robertson}}}, \bibinfo {author}
  {\bibfnamefont {J.}~\bibnamefont {{Sakstein}}},\ and\ \bibinfo {author}
  {\bibfnamefont {J.}~\bibnamefont {{Zavala}}},\ }\bibfield  {title} {\bibinfo
  {title} {{Astrophysical Tests of Dark Matter Self-Interactions}},\ }\href
  {https://doi.org/10.48550/arXiv.2207.10638} {\bibfield  {journal} {\bibinfo
  {journal} {arXiv e-prints}\ ,\ \bibinfo {eid} {arXiv:2207.10638}} (\bibinfo
  {year} {2022})},\ \Eprint {https://arxiv.org/abs/2207.10638}
  {arXiv:2207.10638 [astro-ph.CO]} \BibitemShut {NoStop}%
\bibitem [{\citenamefont {Bullock}\ and\ \citenamefont
  {Boylan-Kolchin}(2017)}]{Bullock:2017xww}%
  \BibitemOpen
  \bibfield  {author} {\bibinfo {author} {\bibfnamefont {J.~S.}\ \bibnamefont
  {Bullock}}\ and\ \bibinfo {author} {\bibfnamefont {M.}~\bibnamefont
  {Boylan-Kolchin}},\ }\bibfield  {title} {\bibinfo {title} {{Small-Scale
  Challenges to the $\Lambda$CDM Paradigm}},\ }\href
  {https://doi.org/10.1146/annurev-astro-091916-055313} {\bibfield  {journal}
  {\bibinfo  {journal} {Ann. Rev. Astron. Astrophys.}\ }\textbf {\bibinfo
  {volume} {55}},\ \bibinfo {pages} {343} (\bibinfo {year} {2017})},\ \Eprint
  {https://arxiv.org/abs/1707.04256} {arXiv:1707.04256 [astro-ph.CO]}
  \BibitemShut {NoStop}%
\bibitem [{\citenamefont {{Plummer}}(1911)}]{1911MNRAS..71..460P}%
  \BibitemOpen
  \bibfield  {author} {\bibinfo {author} {\bibfnamefont {H.~C.}\ \bibnamefont
  {{Plummer}}},\ }\bibfield  {title} {\bibinfo {title} {{On the problem of
  distribution in globular star clusters}},\ }\href
  {https://doi.org/10.1093/mnras/71.5.460} {\bibfield  {journal} {\bibinfo
  {journal} {Mon. Not. Roy. Astron. Soc.}\ }\textbf {\bibinfo {volume} {71}},\
  \bibinfo {pages} {460} (\bibinfo {year} {1911})}\BibitemShut {NoStop}%
\bibitem [{\citenamefont {{Jensen}}\ \emph {et~al.}(2024)\citenamefont
  {{Jensen}}, \citenamefont {{Hayes}}, \citenamefont {{Sestito}}, \citenamefont
  {{McConnachie}}, \citenamefont {{Waller}}, \citenamefont {{Smith}},
  \citenamefont {{Navarro}},\ and\ \citenamefont
  {{Venn}}}]{2024MNRAS.527.4209J}%
  \BibitemOpen
  \bibfield  {author} {\bibinfo {author} {\bibfnamefont {J.}~\bibnamefont
  {{Jensen}}}, \bibinfo {author} {\bibfnamefont {C.~R.}\ \bibnamefont
  {{Hayes}}}, \bibinfo {author} {\bibfnamefont {F.}~\bibnamefont {{Sestito}}},
  \bibinfo {author} {\bibfnamefont {A.~W.}\ \bibnamefont {{McConnachie}}},
  \bibinfo {author} {\bibfnamefont {F.}~\bibnamefont {{Waller}}}, \bibinfo
  {author} {\bibfnamefont {S.~E.~T.}\ \bibnamefont {{Smith}}}, \bibinfo
  {author} {\bibfnamefont {J.}~\bibnamefont {{Navarro}}},\ and\ \bibinfo
  {author} {\bibfnamefont {K.~A.}\ \bibnamefont {{Venn}}},\ }\bibfield  {title}
  {\bibinfo {title} {{Small-scale stellar haloes: detecting low surface
  brightness features in the outskirts of Milky Way dwarf satellites}},\ }\href
  {https://doi.org/10.1093/mnras/stad3322} {\bibfield  {journal} {\bibinfo
  {journal} {Mon. Not. Roy. Astron. Soc.}\ }\textbf {\bibinfo {volume} {527}},\
  \bibinfo {pages} {4209} (\bibinfo {year} {2024})},\ \Eprint
  {https://arxiv.org/abs/2308.07394} {arXiv:2308.07394 [astro-ph.GA]}
  \BibitemShut {NoStop}%
\bibitem [{\citenamefont {{Tau}}\ \emph {et~al.}(2024)\citenamefont {{Tau}},
  \citenamefont {{Vivas}},\ and\ \citenamefont
  {{Mart{\'\i}nez-V{\'a}zquez}}}]{2024AJ....167...57T}%
  \BibitemOpen
  \bibfield  {author} {\bibinfo {author} {\bibfnamefont {E.~A.}\ \bibnamefont
  {{Tau}}}, \bibinfo {author} {\bibfnamefont {A.~K.}\ \bibnamefont {{Vivas}}},\
  and\ \bibinfo {author} {\bibfnamefont {C.~E.}\ \bibnamefont
  {{Mart{\'\i}nez-V{\'a}zquez}}},\ }\bibfield  {title} {\bibinfo {title}
  {{Extended Stellar Populations in Ultrafaint Dwarf Galaxies}},\ }\href
  {https://doi.org/10.3847/1538-3881/ad1509} {\bibfield  {journal} {\bibinfo
  {journal} {The Astronomical Journal}\ }\textbf {\bibinfo {volume} {167}},\
  \bibinfo {eid} {57} (\bibinfo {year} {2024})},\ \Eprint
  {https://arxiv.org/abs/2312.07279} {arXiv:2312.07279 [astro-ph.GA]}
  \BibitemShut {NoStop}%
\bibitem [{\citenamefont {Bruce}\ \emph {et~al.}(2023)\citenamefont {Bruce},
  \citenamefont {Li}, \citenamefont {Pace}, \citenamefont {Heiger},
  \citenamefont {Song},\ and\ \citenamefont {Simon}}]{Bruce_2023}%
  \BibitemOpen
  \bibfield  {author} {\bibinfo {author} {\bibfnamefont {J.}~\bibnamefont
  {Bruce}}, \bibinfo {author} {\bibfnamefont {T.~S.}\ \bibnamefont {Li}},
  \bibinfo {author} {\bibfnamefont {A.~B.}\ \bibnamefont {Pace}}, \bibinfo
  {author} {\bibfnamefont {M.}~\bibnamefont {Heiger}}, \bibinfo {author}
  {\bibfnamefont {Y.-Y.}\ \bibnamefont {Song}},\ and\ \bibinfo {author}
  {\bibfnamefont {J.~D.}\ \bibnamefont {Simon}},\ }\bibfield  {title} {\bibinfo
  {title} {Spectroscopic analysis of milky way outer halo satellites: Aquarius
  ii and boötes ii},\ }\href {https://doi.org/10.3847/1538-4357/acc943}
  {\bibfield  {journal} {\bibinfo  {journal} {The Astrophysical Journal}\
  }\textbf {\bibinfo {volume} {950}},\ \bibinfo {pages} {167} (\bibinfo {year}
  {2023})}\BibitemShut {NoStop}%
\bibitem [{\citenamefont {Chiti}\ \emph {et~al.}(2022)\citenamefont {Chiti},
  \citenamefont {Simon}, \citenamefont {Frebel}, \citenamefont {Pace},
  \citenamefont {Ji},\ and\ \citenamefont {Li}}]{Chiti_2022}%
  \BibitemOpen
  \bibfield  {author} {\bibinfo {author} {\bibfnamefont {A.}~\bibnamefont
  {Chiti}}, \bibinfo {author} {\bibfnamefont {J.~D.}\ \bibnamefont {Simon}},
  \bibinfo {author} {\bibfnamefont {A.}~\bibnamefont {Frebel}}, \bibinfo
  {author} {\bibfnamefont {A.~B.}\ \bibnamefont {Pace}}, \bibinfo {author}
  {\bibfnamefont {A.~P.}\ \bibnamefont {Ji}},\ and\ \bibinfo {author}
  {\bibfnamefont {T.~S.}\ \bibnamefont {Li}},\ }\bibfield  {title} {\bibinfo
  {title} {Magellan/imacs spectroscopy of grus i: A low metallicity ultra-faint
  dwarf galaxy*},\ }\href {https://doi.org/10.3847/1538-4357/ac96ed} {\bibfield
   {journal} {\bibinfo  {journal} {The Astrophysical Journal}\ }\textbf
  {\bibinfo {volume} {939}},\ \bibinfo {pages} {41} (\bibinfo {year}
  {2022})}\BibitemShut {NoStop}%
\bibitem [{\citenamefont {Tan}\ \emph {et~al.}(2025)\citenamefont {Tan},
  \citenamefont {Cerny}, \citenamefont {Drlica-Wagner}, \citenamefont {Pace},
  \citenamefont {Geha}, \citenamefont {Ji}, \citenamefont {Li}, \citenamefont
  {Adamów}, \citenamefont {Anbajagane}, \citenamefont {Bom}, \citenamefont
  {Carballo-Bello}, \citenamefont {Carlin}, \citenamefont {Chang},
  \citenamefont {Chaturvedi}, \citenamefont {Chiti}, \citenamefont {Choi},
  \citenamefont {Collins}, \citenamefont {Doliva-Dolinsky}, \citenamefont
  {Ferguson}, \citenamefont {Gruendl}, \citenamefont {James}, \citenamefont
  {Limberg}, \citenamefont {Navabi}, \citenamefont {Martínez-Delgado},
  \citenamefont {Martínez-Vázquez}, \citenamefont {Medina}, \citenamefont
  {Mutlu-Pakdil}, \citenamefont {Nidever}, \citenamefont {Noël}, \citenamefont
  {Riley}, \citenamefont {Sakowska}, \citenamefont {Sand}, \citenamefont
  {Sharp}, \citenamefont {Stringfellow}, \citenamefont {Tolley}, \citenamefont
  {Tucker}, \citenamefont {Vivas},\ and\ \citenamefont
  {Collaboration)}}]{Tan_2025}%
  \BibitemOpen
  \bibfield  {author} {\bibinfo {author} {\bibfnamefont {C.~Y.}\ \bibnamefont
  {Tan}}, \bibinfo {author} {\bibfnamefont {W.}~\bibnamefont {Cerny}}, \bibinfo
  {author} {\bibfnamefont {A.}~\bibnamefont {Drlica-Wagner}}, \bibinfo {author}
  {\bibfnamefont {A.~B.}\ \bibnamefont {Pace}}, \bibinfo {author}
  {\bibfnamefont {M.}~\bibnamefont {Geha}}, \bibinfo {author} {\bibfnamefont
  {A.~P.}\ \bibnamefont {Ji}}, \bibinfo {author} {\bibfnamefont {T.~S.}\
  \bibnamefont {Li}}, \bibinfo {author} {\bibfnamefont {M.}~\bibnamefont
  {Adamów}}, \bibinfo {author} {\bibfnamefont {D.}~\bibnamefont {Anbajagane}},
  \bibinfo {author} {\bibfnamefont {C.~R.}\ \bibnamefont {Bom}}, \bibinfo
  {author} {\bibfnamefont {J.~A.}\ \bibnamefont {Carballo-Bello}}, \bibinfo
  {author} {\bibfnamefont {J.~L.}\ \bibnamefont {Carlin}}, \bibinfo {author}
  {\bibfnamefont {C.}~\bibnamefont {Chang}}, \bibinfo {author} {\bibfnamefont
  {A.}~\bibnamefont {Chaturvedi}}, \bibinfo {author} {\bibfnamefont
  {A.}~\bibnamefont {Chiti}}, \bibinfo {author} {\bibfnamefont
  {Y.}~\bibnamefont {Choi}}, \bibinfo {author} {\bibfnamefont {M.~L.~M.}\
  \bibnamefont {Collins}}, \bibinfo {author} {\bibfnamefont {A.}~\bibnamefont
  {Doliva-Dolinsky}}, \bibinfo {author} {\bibfnamefont {P.~S.}\ \bibnamefont
  {Ferguson}}, \bibinfo {author} {\bibfnamefont {R.~A.}\ \bibnamefont
  {Gruendl}}, \bibinfo {author} {\bibfnamefont {D.~J.}\ \bibnamefont {James}},
  \bibinfo {author} {\bibfnamefont {G.}~\bibnamefont {Limberg}}, \bibinfo
  {author} {\bibfnamefont {M.}~\bibnamefont {Navabi}}, \bibinfo {author}
  {\bibfnamefont {D.}~\bibnamefont {Martínez-Delgado}}, \bibinfo {author}
  {\bibfnamefont {C.~E.}\ \bibnamefont {Martínez-Vázquez}}, \bibinfo {author}
  {\bibfnamefont {G.~E.}\ \bibnamefont {Medina}}, \bibinfo {author}
  {\bibfnamefont {B.}~\bibnamefont {Mutlu-Pakdil}}, \bibinfo {author}
  {\bibfnamefont {D.~L.}\ \bibnamefont {Nidever}}, \bibinfo {author}
  {\bibfnamefont {N.~E.~D.}\ \bibnamefont {Noël}}, \bibinfo {author}
  {\bibfnamefont {A.~H.}\ \bibnamefont {Riley}}, \bibinfo {author}
  {\bibfnamefont {J.~D.}\ \bibnamefont {Sakowska}}, \bibinfo {author}
  {\bibfnamefont {D.~J.}\ \bibnamefont {Sand}}, \bibinfo {author}
  {\bibfnamefont {J.}~\bibnamefont {Sharp}}, \bibinfo {author} {\bibfnamefont
  {G.~S.}\ \bibnamefont {Stringfellow}}, \bibinfo {author} {\bibfnamefont
  {C.}~\bibnamefont {Tolley}}, \bibinfo {author} {\bibfnamefont {D.~L.}\
  \bibnamefont {Tucker}}, \bibinfo {author} {\bibfnamefont {A.~K.}\
  \bibnamefont {Vivas}},\ and\ \bibinfo {author} {\bibfnamefont
  {D.}~\bibnamefont {Collaboration)}},\ }\bibfield  {title} {\bibinfo {title}
  {A pride of satellites in the constellation leo? discovery of the leo vi
  milky way satellite ultra-faint dwarf galaxy with delve early data release
  3},\ }\href {https://doi.org/10.3847/1538-4357/ad9b0c} {\bibfield  {journal}
  {\bibinfo  {journal} {The Astrophysical Journal}\ }\textbf {\bibinfo {volume}
  {979}},\ \bibinfo {pages} {176} (\bibinfo {year} {2025})}\BibitemShut
  {NoStop}%
\bibitem [{\citenamefont {{Kim}}\ \emph {et~al.}(2016)\citenamefont {{Kim}},
  \citenamefont {{Jerjen}}, \citenamefont {{Geha}}, \citenamefont {{Chiti}},
  \citenamefont {{Milone}}, \citenamefont {{Da Costa}}, \citenamefont
  {{Mackey}}, \citenamefont {{Frebel}},\ and\ \citenamefont
  {{Conn}}}]{2016ApJ...833...16K}%
  \BibitemOpen
  \bibfield  {author} {\bibinfo {author} {\bibfnamefont {D.}~\bibnamefont
  {{Kim}}}, \bibinfo {author} {\bibfnamefont {H.}~\bibnamefont {{Jerjen}}},
  \bibinfo {author} {\bibfnamefont {M.}~\bibnamefont {{Geha}}}, \bibinfo
  {author} {\bibfnamefont {A.}~\bibnamefont {{Chiti}}}, \bibinfo {author}
  {\bibfnamefont {A.~P.}\ \bibnamefont {{Milone}}}, \bibinfo {author}
  {\bibfnamefont {G.}~\bibnamefont {{Da Costa}}}, \bibinfo {author}
  {\bibfnamefont {D.}~\bibnamefont {{Mackey}}}, \bibinfo {author}
  {\bibfnamefont {A.}~\bibnamefont {{Frebel}}},\ and\ \bibinfo {author}
  {\bibfnamefont {B.}~\bibnamefont {{Conn}}},\ }\bibfield  {title} {\bibinfo
  {title} {{Portrait of a Dark Horse: a Photometric and Spectroscopic Study of
  the Ultra-faint Milky Way Satellite Pegasus III}},\ }\href
  {https://doi.org/10.3847/0004-637X/833/1/16} {\bibfield  {journal} {\bibinfo
  {journal} {\apj}\ }\textbf {\bibinfo {volume} {833}},\ \bibinfo {eid} {16}
  (\bibinfo {year} {2016})},\ \Eprint {https://arxiv.org/abs/1608.04934}
  {arXiv:1608.04934 [astro-ph.GA]} \BibitemShut {NoStop}%
\bibitem [{\citenamefont {{Cerny}}\ \emph {et~al.}(2023)\citenamefont
  {{Cerny}}, \citenamefont {{Simon}}, \citenamefont {{Li}}, \citenamefont
  {{Drlica-Wagner}}, \citenamefont {{Pace}}, \citenamefont
  {{Mart{\'\i}nez-V{\'a}zquez}}, \citenamefont {{Riley}}, \citenamefont
  {{Mutlu-Pakdil}}, \citenamefont {{Mau}}, \citenamefont {{Ferguson}},
  \citenamefont {{Erkal}}, \citenamefont {{Munoz}}, \citenamefont {{Bom}},
  \citenamefont {{Carlin}}, \citenamefont {{Carollo}}, \citenamefont {{Choi}},
  \citenamefont {{Ji}}, \citenamefont {{Manwadkar}}, \citenamefont
  {{Mart{\'\i}nez-Delgado}}, \citenamefont {{Miller}}, \citenamefont
  {{No{\"e}l}}, \citenamefont {{Sakowska}}, \citenamefont {{Sand}},
  \citenamefont {{Stringfellow}}, \citenamefont {{Tollerud}}, \citenamefont
  {{Vivas}}, \citenamefont {{Carballo-Bello}}, \citenamefont
  {{Hernandez-Lang}}, \citenamefont {{James}}, \citenamefont {{Nidever}},
  \citenamefont {{Castellon}}, \citenamefont {{Olsen}}, \citenamefont
  {{Zenteno}},\ and\ \citenamefont {{Delve
  Collaboration}}}]{2023ApJ...942..111C}%
  \BibitemOpen
  \bibfield  {author} {\bibinfo {author} {\bibfnamefont {W.}~\bibnamefont
  {{Cerny}}}, \bibinfo {author} {\bibfnamefont {J.~D.}\ \bibnamefont
  {{Simon}}}, \bibinfo {author} {\bibfnamefont {T.~S.}\ \bibnamefont {{Li}}},
  \bibinfo {author} {\bibfnamefont {A.}~\bibnamefont {{Drlica-Wagner}}},
  \bibinfo {author} {\bibfnamefont {A.~B.}\ \bibnamefont {{Pace}}}, \bibinfo
  {author} {\bibfnamefont {C.~E.}\ \bibnamefont {{Mart{\'\i}nez-V{\'a}zquez}}},
  \bibinfo {author} {\bibfnamefont {A.~H.}\ \bibnamefont {{Riley}}}, \bibinfo
  {author} {\bibfnamefont {B.}~\bibnamefont {{Mutlu-Pakdil}}}, \bibinfo
  {author} {\bibfnamefont {S.}~\bibnamefont {{Mau}}}, \bibinfo {author}
  {\bibfnamefont {P.~S.}\ \bibnamefont {{Ferguson}}}, \bibinfo {author}
  {\bibfnamefont {D.}~\bibnamefont {{Erkal}}}, \bibinfo {author} {\bibfnamefont
  {R.~R.}\ \bibnamefont {{Munoz}}}, \bibinfo {author} {\bibfnamefont {C.~R.}\
  \bibnamefont {{Bom}}}, \bibinfo {author} {\bibfnamefont {J.~L.}\ \bibnamefont
  {{Carlin}}}, \bibinfo {author} {\bibfnamefont {D.}~\bibnamefont {{Carollo}}},
  \bibinfo {author} {\bibfnamefont {Y.}~\bibnamefont {{Choi}}}, \bibinfo
  {author} {\bibfnamefont {A.~P.}\ \bibnamefont {{Ji}}}, \bibinfo {author}
  {\bibfnamefont {V.}~\bibnamefont {{Manwadkar}}}, \bibinfo {author}
  {\bibfnamefont {D.}~\bibnamefont {{Mart{\'\i}nez-Delgado}}}, \bibinfo
  {author} {\bibfnamefont {A.~E.}\ \bibnamefont {{Miller}}}, \bibinfo {author}
  {\bibfnamefont {N.~E.~D.}\ \bibnamefont {{No{\"e}l}}}, \bibinfo {author}
  {\bibfnamefont {J.~D.}\ \bibnamefont {{Sakowska}}}, \bibinfo {author}
  {\bibfnamefont {D.~J.}\ \bibnamefont {{Sand}}}, \bibinfo {author}
  {\bibfnamefont {G.~S.}\ \bibnamefont {{Stringfellow}}}, \bibinfo {author}
  {\bibfnamefont {E.~J.}\ \bibnamefont {{Tollerud}}}, \bibinfo {author}
  {\bibfnamefont {A.~K.}\ \bibnamefont {{Vivas}}}, \bibinfo {author}
  {\bibfnamefont {J.~A.}\ \bibnamefont {{Carballo-Bello}}}, \bibinfo {author}
  {\bibfnamefont {D.}~\bibnamefont {{Hernandez-Lang}}}, \bibinfo {author}
  {\bibfnamefont {D.~J.}\ \bibnamefont {{James}}}, \bibinfo {author}
  {\bibfnamefont {D.~L.}\ \bibnamefont {{Nidever}}}, \bibinfo {author}
  {\bibfnamefont {J.~L.~N.}\ \bibnamefont {{Castellon}}}, \bibinfo {author}
  {\bibfnamefont {K.~A.~G.}\ \bibnamefont {{Olsen}}}, \bibinfo {author}
  {\bibfnamefont {A.}~\bibnamefont {{Zenteno}}},\ and\ \bibinfo {author}
  {\bibnamefont {{Delve Collaboration}}},\ }\bibfield  {title} {\bibinfo
  {title} {{Pegasus IV: Discovery and Spectroscopic Confirmation of an
  Ultra-faint Dwarf Galaxy in the Constellation Pegasus}},\ }\href
  {https://doi.org/10.3847/1538-4357/aca1c3} {\bibfield  {journal} {\bibinfo
  {journal} {\apj}\ }\textbf {\bibinfo {volume} {942}},\ \bibinfo {eid} {111}
  (\bibinfo {year} {2023})},\ \Eprint {https://arxiv.org/abs/2203.11788}
  {arXiv:2203.11788 [astro-ph.GA]} \BibitemShut {NoStop}%
\bibitem [{\citenamefont {{Kirby}}\ \emph {et~al.}(2015)\citenamefont
  {{Kirby}}, \citenamefont {{Simon}},\ and\ \citenamefont
  {{Cohen}}}]{2015ApJ...810...56K}%
  \BibitemOpen
  \bibfield  {author} {\bibinfo {author} {\bibfnamefont {E.~N.}\ \bibnamefont
  {{Kirby}}}, \bibinfo {author} {\bibfnamefont {J.~D.}\ \bibnamefont
  {{Simon}}},\ and\ \bibinfo {author} {\bibfnamefont {J.~G.}\ \bibnamefont
  {{Cohen}}},\ }\bibfield  {title} {\bibinfo {title} {{Spectroscopic
  Confirmation of the Dwarf Galaxies Hydra II and Pisces II and the Globular
  Cluster Laevens 1}},\ }\href {https://doi.org/10.1088/0004-637X/810/1/56}
  {\bibfield  {journal} {\bibinfo  {journal} {\apj}\ }\textbf {\bibinfo
  {volume} {810}},\ \bibinfo {eid} {56} (\bibinfo {year} {2015})},\ \Eprint
  {https://arxiv.org/abs/1506.01021} {arXiv:1506.01021 [astro-ph.GA]}
  \BibitemShut {NoStop}%
\bibitem [{\citenamefont {{Koposov}}\ \emph {et~al.}(2015)\citenamefont
  {{Koposov}}, \citenamefont {{Casey}}, \citenamefont {{Belokurov}},
  \citenamefont {{Lewis}}, \citenamefont {{Gilmore}}, \citenamefont {{Worley}},
  \citenamefont {{Hourihane}}, \citenamefont {{Randich}}, \citenamefont
  {{Bensby}}, \citenamefont {{Bragaglia}}, \citenamefont {{Bergemann}},
  \citenamefont {{Carraro}}, \citenamefont {{Costado}}, \citenamefont
  {{Flaccomio}}, \citenamefont {{Francois}}, \citenamefont {{Heiter}},
  \citenamefont {{Hill}}, \citenamefont {{Jofre}}, \citenamefont {{Lando}},
  \citenamefont {{Lanzafame}}, \citenamefont {{de Laverny}}, \citenamefont
  {{Monaco}}, \citenamefont {{Morbidelli}}, \citenamefont {{Sbordone}},
  \citenamefont {{Mikolaitis}},\ and\ \citenamefont
  {{Ryde}}}]{2015ApJ...811...62K}%
  \BibitemOpen
  \bibfield  {author} {\bibinfo {author} {\bibfnamefont {S.~E.}\ \bibnamefont
  {{Koposov}}}, \bibinfo {author} {\bibfnamefont {A.~R.}\ \bibnamefont
  {{Casey}}}, \bibinfo {author} {\bibfnamefont {V.}~\bibnamefont
  {{Belokurov}}}, \bibinfo {author} {\bibfnamefont {J.~R.}\ \bibnamefont
  {{Lewis}}}, \bibinfo {author} {\bibfnamefont {G.}~\bibnamefont {{Gilmore}}},
  \bibinfo {author} {\bibfnamefont {C.}~\bibnamefont {{Worley}}}, \bibinfo
  {author} {\bibfnamefont {A.}~\bibnamefont {{Hourihane}}}, \bibinfo {author}
  {\bibfnamefont {S.}~\bibnamefont {{Randich}}}, \bibinfo {author}
  {\bibfnamefont {T.}~\bibnamefont {{Bensby}}}, \bibinfo {author}
  {\bibfnamefont {A.}~\bibnamefont {{Bragaglia}}}, \bibinfo {author}
  {\bibfnamefont {M.}~\bibnamefont {{Bergemann}}}, \bibinfo {author}
  {\bibfnamefont {G.}~\bibnamefont {{Carraro}}}, \bibinfo {author}
  {\bibfnamefont {M.~T.}\ \bibnamefont {{Costado}}}, \bibinfo {author}
  {\bibfnamefont {E.}~\bibnamefont {{Flaccomio}}}, \bibinfo {author}
  {\bibfnamefont {P.}~\bibnamefont {{Francois}}}, \bibinfo {author}
  {\bibfnamefont {U.}~\bibnamefont {{Heiter}}}, \bibinfo {author}
  {\bibfnamefont {V.}~\bibnamefont {{Hill}}}, \bibinfo {author} {\bibfnamefont
  {P.}~\bibnamefont {{Jofre}}}, \bibinfo {author} {\bibfnamefont
  {C.}~\bibnamefont {{Lando}}}, \bibinfo {author} {\bibfnamefont {A.~C.}\
  \bibnamefont {{Lanzafame}}}, \bibinfo {author} {\bibfnamefont
  {P.}~\bibnamefont {{de Laverny}}}, \bibinfo {author} {\bibfnamefont
  {L.}~\bibnamefont {{Monaco}}}, \bibinfo {author} {\bibfnamefont
  {L.}~\bibnamefont {{Morbidelli}}}, \bibinfo {author} {\bibfnamefont
  {L.}~\bibnamefont {{Sbordone}}}, \bibinfo {author} {\bibfnamefont
  {{\v{S}}.}~\bibnamefont {{Mikolaitis}}},\ and\ \bibinfo {author}
  {\bibfnamefont {N.}~\bibnamefont {{Ryde}}},\ }\bibfield  {title} {\bibinfo
  {title} {{Kinematics and Chemistry of Recently Discovered Reticulum 2 and
  Horologium 1 Dwarf Galaxies}},\ }\href
  {https://doi.org/10.1088/0004-637X/811/1/62} {\bibfield  {journal} {\bibinfo
  {journal} {\apj}\ }\textbf {\bibinfo {volume} {811}},\ \bibinfo {eid} {62}
  (\bibinfo {year} {2015})},\ \Eprint {https://arxiv.org/abs/1504.07916}
  {arXiv:1504.07916 [astro-ph.GA]} \BibitemShut {NoStop}%
\bibitem [{\citenamefont {Walker}\ \emph {et~al.}(2016)\citenamefont {Walker},
  \citenamefont {Mateo}, \citenamefont {Olszewski}, \citenamefont {Koposov},
  \citenamefont {Belokurov}, \citenamefont {Jethwa}, \citenamefont {Nidever},
  \citenamefont {Bonnivard}, \citenamefont {III}, \citenamefont {Bell},\ and\
  \citenamefont {Loebman}}]{Walker_2016}%
  \BibitemOpen
  \bibfield  {author} {\bibinfo {author} {\bibfnamefont {M.~G.}\ \bibnamefont
  {Walker}}, \bibinfo {author} {\bibfnamefont {M.}~\bibnamefont {Mateo}},
  \bibinfo {author} {\bibfnamefont {E.~W.}\ \bibnamefont {Olszewski}}, \bibinfo
  {author} {\bibfnamefont {S.}~\bibnamefont {Koposov}}, \bibinfo {author}
  {\bibfnamefont {V.}~\bibnamefont {Belokurov}}, \bibinfo {author}
  {\bibfnamefont {P.}~\bibnamefont {Jethwa}}, \bibinfo {author} {\bibfnamefont
  {D.~L.}\ \bibnamefont {Nidever}}, \bibinfo {author} {\bibfnamefont
  {V.}~\bibnamefont {Bonnivard}}, \bibinfo {author} {\bibfnamefont {J.~I.~B.}\
  \bibnamefont {III}}, \bibinfo {author} {\bibfnamefont {E.~F.}\ \bibnamefont
  {Bell}},\ and\ \bibinfo {author} {\bibfnamefont {S.~R.}\ \bibnamefont
  {Loebman}},\ }\bibfield  {title} {\bibinfo {title} {Magellan/m2fs
  spectroscopy of tucana 2 and grus 1*},\ }\href
  {https://doi.org/10.3847/0004-637X/819/1/53} {\bibfield  {journal} {\bibinfo
  {journal} {The Astrophysical Journal}\ }\textbf {\bibinfo {volume} {819}},\
  \bibinfo {pages} {53} (\bibinfo {year} {2016})}\BibitemShut {NoStop}%
\bibitem [{\citenamefont {Pace}(2024)}]{Pace:2024sys}%
  \BibitemOpen
  \bibfield  {author} {\bibinfo {author} {\bibfnamefont {A.~B.}\ \bibnamefont
  {Pace}},\ }\bibfield  {title} {\bibinfo {title} {{The Local Volume Database:
  a library of the observed properties of nearby dwarf galaxies and star
  clusters}},\ }\href@noop {} {\  (\bibinfo {year} {2024})},\ \Eprint
  {https://arxiv.org/abs/2411.07424} {arXiv:2411.07424 [astro-ph.GA]}
  \BibitemShut {NoStop}%
\bibitem [{\citenamefont {Wolf}\ \emph {et~al.}(2010)\citenamefont {Wolf},
  \citenamefont {Martinez}, \citenamefont {Bullock}, \citenamefont
  {Kaplinghat}, \citenamefont {Geha}, \citenamefont {Munoz}, \citenamefont
  {Simon},\ and\ \citenamefont {Avedo}}]{Wolf:2009tu}%
  \BibitemOpen
  \bibfield  {author} {\bibinfo {author} {\bibfnamefont {J.}~\bibnamefont
  {Wolf}}, \bibinfo {author} {\bibfnamefont {G.~D.}\ \bibnamefont {Martinez}},
  \bibinfo {author} {\bibfnamefont {J.~S.}\ \bibnamefont {Bullock}}, \bibinfo
  {author} {\bibfnamefont {M.}~\bibnamefont {Kaplinghat}}, \bibinfo {author}
  {\bibfnamefont {M.}~\bibnamefont {Geha}}, \bibinfo {author} {\bibfnamefont
  {R.~R.}\ \bibnamefont {Munoz}}, \bibinfo {author} {\bibfnamefont {J.~D.}\
  \bibnamefont {Simon}},\ and\ \bibinfo {author} {\bibfnamefont {F.~F.}\
  \bibnamefont {Avedo}},\ }\bibfield  {title} {\bibinfo {title} {{Accurate
  Masses for Dispersion-supported Galaxies}},\ }\href
  {https://doi.org/10.1111/j.1365-2966.2010.16753.x} {\bibfield  {journal}
  {\bibinfo  {journal} {Mon. Not. Roy. Astron. Soc.}\ }\textbf {\bibinfo
  {volume} {406}},\ \bibinfo {pages} {1220} (\bibinfo {year} {2010})},\ \Eprint
  {https://arxiv.org/abs/0908.2995} {arXiv:0908.2995 [astro-ph.CO]}
  \BibitemShut {NoStop}%
\bibitem [{\citenamefont {{Dejonghe}}(1987)}]{1987MNRAS.224...13D}%
  \BibitemOpen
  \bibfield  {author} {\bibinfo {author} {\bibfnamefont {H.}~\bibnamefont
  {{Dejonghe}}},\ }\bibfield  {title} {\bibinfo {title} {{A completely
  analytical family of anisotropic Plummer models}},\ }\href
  {https://doi.org/10.1093/mnras/224.1.13} {\bibfield  {journal} {\bibinfo
  {journal} {\mnras}\ }\textbf {\bibinfo {volume} {224}},\ \bibinfo {pages}
  {13} (\bibinfo {year} {1987})}\BibitemShut {NoStop}%
\bibitem [{\citenamefont {{Merritt}}(1987)}]{1987ApJ...313..121M}%
  \BibitemOpen
  \bibfield  {author} {\bibinfo {author} {\bibfnamefont {D.}~\bibnamefont
  {{Merritt}}},\ }\bibfield  {title} {\bibinfo {title} {{The Distribution of
  Dark Matter in the Coma Cluster}},\ }\href {https://doi.org/10.1086/164953}
  {\bibfield  {journal} {\bibinfo  {journal} {\apj}\ }\textbf {\bibinfo
  {volume} {313}},\ \bibinfo {pages} {121} (\bibinfo {year}
  {1987})}\BibitemShut {NoStop}%
\bibitem [{\citenamefont {{Evans}}\ \emph {et~al.}(2009)\citenamefont
  {{Evans}}, \citenamefont {{An}},\ and\ \citenamefont
  {{Walker}}}]{2009MNRAS.393L..50E}%
  \BibitemOpen
  \bibfield  {author} {\bibinfo {author} {\bibfnamefont {N.~W.}\ \bibnamefont
  {{Evans}}}, \bibinfo {author} {\bibfnamefont {J.}~\bibnamefont {{An}}},\ and\
  \bibinfo {author} {\bibfnamefont {M.~G.}\ \bibnamefont {{Walker}}},\
  }\bibfield  {title} {\bibinfo {title} {{Cores and cusps in the dwarf
  spheroidals}},\ }\href {https://doi.org/10.1111/j.1745-3933.2008.00596.x}
  {\bibfield  {journal} {\bibinfo  {journal} {\mnras}\ }\textbf {\bibinfo
  {volume} {393}},\ \bibinfo {pages} {L50} (\bibinfo {year} {2009})},\ \Eprint
  {https://arxiv.org/abs/0811.1488} {arXiv:0811.1488 [astro-ph]} \BibitemShut
  {NoStop}%
\bibitem [{\citenamefont {Richardson}\ and\ \citenamefont
  {Fairbairn}(2013)}]{Richardson:2012ig}%
  \BibitemOpen
  \bibfield  {author} {\bibinfo {author} {\bibfnamefont {T.}~\bibnamefont
  {Richardson}}\ and\ \bibinfo {author} {\bibfnamefont {M.}~\bibnamefont
  {Fairbairn}},\ }\bibfield  {title} {\bibinfo {title} {{Analytical Solutions
  to the Mass-Anisotropy Degeneracy with Higher Order Jeans Analysis: A General
  Method}},\ }\href {https://doi.org/10.1093/mnras/stt686} {\bibfield
  {journal} {\bibinfo  {journal} {Mon. Not. Roy. Astron. Soc.}\ }\textbf
  {\bibinfo {volume} {432}},\ \bibinfo {pages} {3361} (\bibinfo {year}
  {2013})},\ \Eprint {https://arxiv.org/abs/1207.1709} {arXiv:1207.1709
  [astro-ph.CO]} \BibitemShut {NoStop}%
\bibitem [{\citenamefont {Diakogiannis}\ \emph {et~al.}(2014)\citenamefont
  {Diakogiannis}, \citenamefont {Lewis},\ and\ \citenamefont
  {Ibata}}]{Diakogiannis_2014}%
  \BibitemOpen
  \bibfield  {author} {\bibinfo {author} {\bibfnamefont {F.~I.}\ \bibnamefont
  {Diakogiannis}}, \bibinfo {author} {\bibfnamefont {G.~F.}\ \bibnamefont
  {Lewis}},\ and\ \bibinfo {author} {\bibfnamefont {R.~A.}\ \bibnamefont
  {Ibata}},\ }\bibfield  {title} {\bibinfo {title} {Resolving the
  mass–anisotropy degeneracy of the spherically symmetric jeans equation –
  i. theoretical foundation},\ }\href {https://doi.org/10.1093/mnras/stu1153}
  {\bibfield  {journal} {\bibinfo  {journal} {Monthly Notices of the Royal
  Astronomical Society}\ }\textbf {\bibinfo {volume} {443}},\ \bibinfo {pages}
  {598–609} (\bibinfo {year} {2014})}\BibitemShut {NoStop}%
\bibitem [{\citenamefont {Boddy}\ \emph {et~al.}(2017)\citenamefont {Boddy},
  \citenamefont {Kumar}, \citenamefont {Strigari},\ and\ \citenamefont
  {Wang}}]{Boddy:2017vpe}%
  \BibitemOpen
  \bibfield  {author} {\bibinfo {author} {\bibfnamefont {K.~K.}\ \bibnamefont
  {Boddy}}, \bibinfo {author} {\bibfnamefont {J.}~\bibnamefont {Kumar}},
  \bibinfo {author} {\bibfnamefont {L.~E.}\ \bibnamefont {Strigari}},\ and\
  \bibinfo {author} {\bibfnamefont {M.-Y.}\ \bibnamefont {Wang}},\ }\bibfield
  {title} {\bibinfo {title} {{Sommerfeld-Enhanced $J$-Factors For Dwarf
  Spheroidal Galaxies}},\ }\href {https://doi.org/10.1103/PhysRevD.95.123008}
  {\bibfield  {journal} {\bibinfo  {journal} {Phys. Rev. D}\ }\textbf {\bibinfo
  {volume} {95}},\ \bibinfo {pages} {123008} (\bibinfo {year} {2017})},\
  \Eprint {https://arxiv.org/abs/1702.00408} {arXiv:1702.00408 [astro-ph.CO]}
  \BibitemShut {NoStop}%
\bibitem [{\citenamefont {Boddy}\ \emph {et~al.}(2019)\citenamefont {Boddy},
  \citenamefont {Kumar}, \citenamefont {Runburg},\ and\ \citenamefont
  {Strigari}}]{Boddy:2019wfg}%
  \BibitemOpen
  \bibfield  {author} {\bibinfo {author} {\bibfnamefont {K.~K.}\ \bibnamefont
  {Boddy}}, \bibinfo {author} {\bibfnamefont {J.}~\bibnamefont {Kumar}},
  \bibinfo {author} {\bibfnamefont {J.}~\bibnamefont {Runburg}},\ and\ \bibinfo
  {author} {\bibfnamefont {L.~E.}\ \bibnamefont {Strigari}},\ }\bibfield
  {title} {\bibinfo {title} {{Angular distribution of gamma-ray emission from
  velocity-dependent dark matter annihilation in subhalos}},\ }\href
  {https://doi.org/10.1103/PhysRevD.100.063019} {\bibfield  {journal} {\bibinfo
   {journal} {Phys. Rev. D}\ }\textbf {\bibinfo {volume} {100}},\ \bibinfo
  {pages} {063019} (\bibinfo {year} {2019})},\ \Eprint
  {https://arxiv.org/abs/1905.03431} {arXiv:1905.03431 [astro-ph.CO]}
  \BibitemShut {NoStop}%
\bibitem [{\citenamefont {Vienneau}\ \emph {et~al.}(2024)\citenamefont
  {Vienneau}, \citenamefont {Evans}, \citenamefont {Hartl}, \citenamefont
  {Bozorgnia}, \citenamefont {Strigari}, \citenamefont {Riley},\ and\
  \citenamefont {Shipp}}]{Vienneau:2024xie}%
  \BibitemOpen
  \bibfield  {author} {\bibinfo {author} {\bibfnamefont {E.}~\bibnamefont
  {Vienneau}}, \bibinfo {author} {\bibfnamefont {A.~J.}\ \bibnamefont {Evans}},
  \bibinfo {author} {\bibfnamefont {O.~V.}\ \bibnamefont {Hartl}}, \bibinfo
  {author} {\bibfnamefont {N.}~\bibnamefont {Bozorgnia}}, \bibinfo {author}
  {\bibfnamefont {L.~E.}\ \bibnamefont {Strigari}}, \bibinfo {author}
  {\bibfnamefont {A.~H.}\ \bibnamefont {Riley}},\ and\ \bibinfo {author}
  {\bibfnamefont {N.}~\bibnamefont {Shipp}},\ }\bibfield  {title} {\bibinfo
  {title} {{Significant impact of Galactic dark matter particles on
  annihilation signals from Sagittarius analogues}},\ }\href
  {https://doi.org/10.1088/1475-7516/2024/10/019} {\bibfield  {journal}
  {\bibinfo  {journal} {JCAP}\ }\textbf {\bibinfo {volume} {10}},\ \bibinfo
  {pages} {019}},\ \Eprint {https://arxiv.org/abs/2403.15544} {arXiv:2403.15544
  [astro-ph.HE]} \BibitemShut {NoStop}%
\bibitem [{\citenamefont {Evans}\ \emph {et~al.}(2016)\citenamefont {Evans},
  \citenamefont {Sanders},\ and\ \citenamefont
  {Geringer-Sameth}}]{Evans:2016xwx}%
  \BibitemOpen
  \bibfield  {author} {\bibinfo {author} {\bibfnamefont {N.~W.}\ \bibnamefont
  {Evans}}, \bibinfo {author} {\bibfnamefont {J.~L.}\ \bibnamefont {Sanders}},\
  and\ \bibinfo {author} {\bibfnamefont {A.}~\bibnamefont {Geringer-Sameth}},\
  }\bibfield  {title} {\bibinfo {title} {{Simple J-Factors and D-Factors for
  Indirect Dark Matter Detection}},\ }\href
  {https://doi.org/10.1103/PhysRevD.93.103512} {\bibfield  {journal} {\bibinfo
  {journal} {Phys. Rev. D}\ }\textbf {\bibinfo {volume} {93}},\ \bibinfo
  {pages} {103512} (\bibinfo {year} {2016})},\ \Eprint
  {https://arxiv.org/abs/1604.05599} {arXiv:1604.05599 [astro-ph.GA]}
  \BibitemShut {NoStop}%
\bibitem [{\citenamefont {Pace}\ and\ \citenamefont
  {Strigari}(2019)}]{Pace:2018tin}%
  \BibitemOpen
  \bibfield  {author} {\bibinfo {author} {\bibfnamefont {A.~B.}\ \bibnamefont
  {Pace}}\ and\ \bibinfo {author} {\bibfnamefont {L.~E.}\ \bibnamefont
  {Strigari}},\ }\bibfield  {title} {\bibinfo {title} {{Scaling Relations for
  Dark Matter Annihilation and Decay Profiles in Dwarf Spheroidal Galaxies}},\
  }\href {https://doi.org/10.1093/mnras/sty2839} {\bibfield  {journal}
  {\bibinfo  {journal} {Mon. Not. Roy. Astron. Soc.}\ }\textbf {\bibinfo
  {volume} {482}},\ \bibinfo {pages} {3480} (\bibinfo {year} {2019})},\ \Eprint
  {https://arxiv.org/abs/1802.06811} {arXiv:1802.06811 [astro-ph.GA]}
  \BibitemShut {NoStop}%
\bibitem [{\citenamefont {Hoskinson}\ \emph {et~al.}(2025)\citenamefont
  {Hoskinson}, \citenamefont {Kumar},\ and\ \citenamefont
  {Sandick}}]{Hoskinson:2024hpk}%
  \BibitemOpen
  \bibfield  {author} {\bibinfo {author} {\bibfnamefont {C.}~\bibnamefont
  {Hoskinson}}, \bibinfo {author} {\bibfnamefont {J.}~\bibnamefont {Kumar}},\
  and\ \bibinfo {author} {\bibfnamefont {P.}~\bibnamefont {Sandick}},\
  }\bibfield  {title} {\bibinfo {title} {{Tools for probing new physics with
  newly discovered gamma-ray targets}},\ }\href
  {https://doi.org/10.1103/PhysRevD.111.063078} {\bibfield  {journal} {\bibinfo
   {journal} {Phys. Rev. D}\ }\textbf {\bibinfo {volume} {111}},\ \bibinfo
  {pages} {063078} (\bibinfo {year} {2025})},\ \Eprint
  {https://arxiv.org/abs/2408.04611} {arXiv:2408.04611 [astro-ph.HE]}
  \BibitemShut {NoStop}%
\bibitem [{\citenamefont {Eddington}(1916)}]{10.1093/mnras/76.7.572}%
  \BibitemOpen
  \bibfield  {author} {\bibinfo {author} {\bibfnamefont {A.~S.}\ \bibnamefont
  {Eddington}},\ }\bibfield  {title} {\bibinfo {title} {The distribution of
  stars in globular clusters},\ }\href {https://doi.org/10.1093/mnras/76.7.572}
  {\bibfield  {journal} {\bibinfo  {journal} {Monthly Notices of the Royal
  Astronomical Society}\ }\textbf {\bibinfo {volume} {76}},\ \bibinfo {pages}
  {572} (\bibinfo {year} {1916})},\ \Eprint
  {https://arxiv.org/abs/https://academic.oup.com/mnras/article-pdf/76/7/572/3902739/mnras76-0572.pdf}
  {https://academic.oup.com/mnras/article-pdf/76/7/572/3902739/mnras76-0572.pdf}
  \BibitemShut {NoStop}%
\bibitem [{\citenamefont {Boucher}\ \emph {et~al.}(2022)\citenamefont
  {Boucher}, \citenamefont {Kumar}, \citenamefont {Le},\ and\ \citenamefont
  {Runburg}}]{Boucher:2021mii}%
  \BibitemOpen
  \bibfield  {author} {\bibinfo {author} {\bibfnamefont {B.}~\bibnamefont
  {Boucher}}, \bibinfo {author} {\bibfnamefont {J.}~\bibnamefont {Kumar}},
  \bibinfo {author} {\bibfnamefont {V.~B.}\ \bibnamefont {Le}},\ and\ \bibinfo
  {author} {\bibfnamefont {J.}~\bibnamefont {Runburg}},\ }\bibfield  {title}
  {\bibinfo {title} {{J-factors for velocity-dependent dark matter
  annihilation}},\ }\href {https://doi.org/10.1103/PhysRevD.106.023025}
  {\bibfield  {journal} {\bibinfo  {journal} {Phys. Rev. D}\ }\textbf {\bibinfo
  {volume} {106}},\ \bibinfo {pages} {023025} (\bibinfo {year} {2022})},\
  \Eprint {https://arxiv.org/abs/2110.09653} {arXiv:2110.09653 [hep-ph]}
  \BibitemShut {NoStop}%
\bibitem [{\citenamefont {{Cerny}}\ \emph {et~al.}(2026)\citenamefont
  {{Cerny}}, \citenamefont {{Bissonette}}, \citenamefont {{Ji}}, \citenamefont
  {{Geha}}, \citenamefont {{Chiti}}, \citenamefont {{Smith}}, \citenamefont
  {{Simon}}, \citenamefont {{Pace}}, \citenamefont {{Kirby}}, \citenamefont
  {{Venn}},\ and\ \citenamefont {et~al.}}]{2026ApJ...999L...8C}%
  \BibitemOpen
  \bibfield  {author} {\bibinfo {author} {\bibfnamefont {W.}~\bibnamefont
  {{Cerny}}}, \bibinfo {author} {\bibfnamefont {D.}~\bibnamefont
  {{Bissonette}}}, \bibinfo {author} {\bibfnamefont {A.~P.}\ \bibnamefont
  {{Ji}}}, \bibinfo {author} {\bibfnamefont {M.}~\bibnamefont {{Geha}}},
  \bibinfo {author} {\bibfnamefont {A.}~\bibnamefont {{Chiti}}}, \bibinfo
  {author} {\bibfnamefont {S.~E.~T.}\ \bibnamefont {{Smith}}}, \bibinfo
  {author} {\bibfnamefont {J.~D.}\ \bibnamefont {{Simon}}}, \bibinfo {author}
  {\bibfnamefont {A.~B.}\ \bibnamefont {{Pace}}}, \bibinfo {author}
  {\bibfnamefont {E.~N.}\ \bibnamefont {{Kirby}}}, \bibinfo {author}
  {\bibfnamefont {K.~A.}\ \bibnamefont {{Venn}}},\ and\ \bibinfo {author}
  {\bibnamefont {et~al.}},\ }\bibfield  {title} {\bibinfo {title} {{No
  Observational Evidence for Dark Matter Nor a Large Metallicity Spread in the
  Extreme Milky Way Satellite Ursa Major III/UNIONS 1}},\ }\href
  {https://doi.org/10.3847/2041-8213/ae29b8} {\bibfield  {journal} {\bibinfo
  {journal} {\apjl}\ }\textbf {\bibinfo {volume} {999}},\ \bibinfo {eid} {L8}
  (\bibinfo {year} {2026})},\ \Eprint {https://arxiv.org/abs/2510.02431}
  {arXiv:2510.02431 [astro-ph.GA]} \BibitemShut {NoStop}%
\bibitem [{\citenamefont {Feroz}\ \emph {et~al.}(2009)\citenamefont {Feroz},
  \citenamefont {Hobson},\ and\ \citenamefont {Bridges}}]{MultiNestDocs}%
  \BibitemOpen
  \bibfield  {author} {\bibinfo {author} {\bibfnamefont {F.}~\bibnamefont
  {Feroz}}, \bibinfo {author} {\bibfnamefont {M.~P.}\ \bibnamefont {Hobson}},\
  and\ \bibinfo {author} {\bibfnamefont {M.}~\bibnamefont {Bridges}},\
  }\bibfield  {title} {\bibinfo {title} {{MultiNest: an efficient and robust
  Bayesian inference tool for cosmology and particle physics}},\ }\href
  {https://doi.org/10.1111/j.1365-2966.2009.14548.x} {\bibfield  {journal}
  {\bibinfo  {journal} {Monthly Notices of the Royal Astronomical Society}\
  }\textbf {\bibinfo {volume} {398}},\ \bibinfo {pages} {1601} (\bibinfo {year}
  {2009})},\ \Eprint
  {https://arxiv.org/abs/https://academic.oup.com/mnras/article-pdf/398/4/1601/3039078/mnras0398-1601.pdf}
  {https://academic.oup.com/mnras/article-pdf/398/4/1601/3039078/mnras0398-1601.pdf}
  \BibitemShut {NoStop}%
\bibitem [{\citenamefont {{Buchner}}\ \emph {et~al.}(2014)\citenamefont
  {{Buchner}}, \citenamefont {{Georgakakis}}, \citenamefont {{Nandra}},
  \citenamefont {{Hsu}}, \citenamefont {{Rangel}}, \citenamefont {{Brightman}},
  \citenamefont {{Merloni}}, \citenamefont {{Salvato}}, \citenamefont
  {{Donley}},\ and\ \citenamefont {{Kocevski}}}]{pymultinestDoc}%
  \BibitemOpen
  \bibfield  {author} {\bibinfo {author} {\bibfnamefont {J.}~\bibnamefont
  {{Buchner}}}, \bibinfo {author} {\bibfnamefont {A.}~\bibnamefont
  {{Georgakakis}}}, \bibinfo {author} {\bibfnamefont {K.}~\bibnamefont
  {{Nandra}}}, \bibinfo {author} {\bibfnamefont {L.}~\bibnamefont {{Hsu}}},
  \bibinfo {author} {\bibfnamefont {C.}~\bibnamefont {{Rangel}}}, \bibinfo
  {author} {\bibfnamefont {M.}~\bibnamefont {{Brightman}}}, \bibinfo {author}
  {\bibfnamefont {A.}~\bibnamefont {{Merloni}}}, \bibinfo {author}
  {\bibfnamefont {M.}~\bibnamefont {{Salvato}}}, \bibinfo {author}
  {\bibfnamefont {J.}~\bibnamefont {{Donley}}},\ and\ \bibinfo {author}
  {\bibfnamefont {D.}~\bibnamefont {{Kocevski}}},\ }\bibfield  {title}
  {\bibinfo {title} {{X-ray spectral modelling of the AGN obscuring region in
  the CDFS: Bayesian model selection and catalogue}},\ }\href
  {https://doi.org/10.1051/0004-6361/201322971} {\bibfield  {journal} {\bibinfo
   {journal} {Astronomy and Astrophysics}\ }\textbf {\bibinfo {volume} {564}},\
  \bibinfo {eid} {A125} (\bibinfo {year} {2014})},\ \Eprint
  {https://arxiv.org/abs/1402.0004} {arXiv:1402.0004 [astro-ph.HE]}
  \BibitemShut {NoStop}%
\end{thebibliography}%



\appendix
\section{Jeans Analysis}
\label{app:jeansdetails}

We utilize the Bayesian inference tool {\tt MultiNest} through the python interface {\tt PyMultiNest} to estimate the posterior \citep{MultiNestDocs,pymultinestDoc}.
We sample over the following flat priors in logarithmic space for velocity anisotropy, halo mass, and concentration parameters:
\begin{equation}\label{eqn:priorsexpanded}
\begin{aligned}
 -1 \leq \log_{10} (1-\beta_\star) \leq +1, \\
 \log_{10}(2 \times 10^{5}) \leq \log_{10} (M_{200} / M_\odot) \leq \log_{10}(7 \times 10^{11}) , \\
 \log_{10}(1.25) \leq \log_{10} (c_{200}) \leq \log_{10}(125) .
\end{aligned}
\end{equation}
For each dSph, we assign a log-normal prior to the projected half-light radius $r_h$ using the value $R_{1/2}$ reported in Ref.~\cite{Pace:2024sys}. We take the reported central value to be the median of the prior $\mu= \ln(R_{1/2})$ and estimate its logarithmic width from the asymmetric uncertainty interval defined as $\nu= \frac{1}{2} \ln{\left( \frac{R_{1/2} + \sigma_+}{R_{1/2} - \sigma_-} \right)}$. MultiNest supplies a unit-hypercube coordinate $u\in(0,1)$ to the inverse standard-normal cumulative distribution function $\Phi^{-1}$ to calculate a strictly positive prior for $r_h$.

\bea
z &=& \Phi^{-1}(u) \sim \mathcal{N}(0,1) ,
\nonumber\\
\psi &=& \mu + \nu z = \mu + \nu \Phi^{-1}(u) \sim \mathcal{N}(\mu,\nu^2) ,
\nonumber\\
r_h &=& \exp (\psi) = \exp [\mu + \nu \Phi^{-1}(u) ] \sim \operatorname{LogNormal}(\mu, \nu^2) .
\eea

Here, $M_{200}$ is the dark matter mass enclosed within the radius
$r_{200}$, where $r_{200}$ is defined as the radius at which the average density of the enclosed dark matter is
200 times the critical density $\rho_c = 3 H_0^2 / 8\pi G_N$ (we take $h=0.7$).
We also define $c_{200} \equiv r_{200} / r_s$.  For an NFW profile, the halo parameters $\rho_s$ and $r_s$ can be expressed in
terms of $M_{200}$ and $c_{200}$, yielding
\bea
\rho_s &=& 9.27 \times 10^{10} h^2 \frac{M_\odot}{\mpc^3}
\left[ \frac{200~ c_{200}^3 }{\ln [1+c_{200}] -  \frac{c_{200}}{1+c_{200} } }  \right] ,
\nonumber\\
\left( \frac{r_s}{\mpc} \right)^3 &=& 8.59 \times 10^{-13} h^{-2}
\left[ \frac{M_{200}}{M_\odot} \frac{c_{200}^{-3}}{200} \right] .
\eea

\end{document}